\documentclass[fleqn,usenatbib]{mnras}

\usepackage{newtxtext,newtxmath}

\usepackage[T1]{fontenc}

\DeclareRobustCommand{\VAN}[3]{#2}
\let\VANthebibliography\thebibliography
\def\thebibliography{\DeclareRobustCommand{\VAN}[3]{##3}\VANthebibliography}

\usepackage{graphicx}	
\usepackage{amsmath}	

\usepackage{scalerel}
\usepackage{lscape}

\newcommand\hi{H\protect\scaleto{$I$}{1.2ex}}

\newcommand\htwoVERO{H$_2$}
\newcommand\htwo{CO}
\newcommand{\kms}{$\,$km$\,$s$^{-1}$}

\newcommand\mJb{mJy beam$^{-1}$}

\newcommand\mhi{$M_{\text{H\protect\scaleto{$I$}{1.2ex}}}$}

\newcommand\mst{$M_\star$}

\usepackage{caption}
\usepackage{subcaption} 
\usepackage{xcolor}
\usepackage{booktabs}
\usepackage{graphbox,graphicx}
\usepackage{txfonts}
\usepackage{hyperref}
\usepackage[toc,page]{appendix}

\newcommand{\placefigopt}{
\begin{figure}
    \centering
    \includegraphics[width=\linewidth]{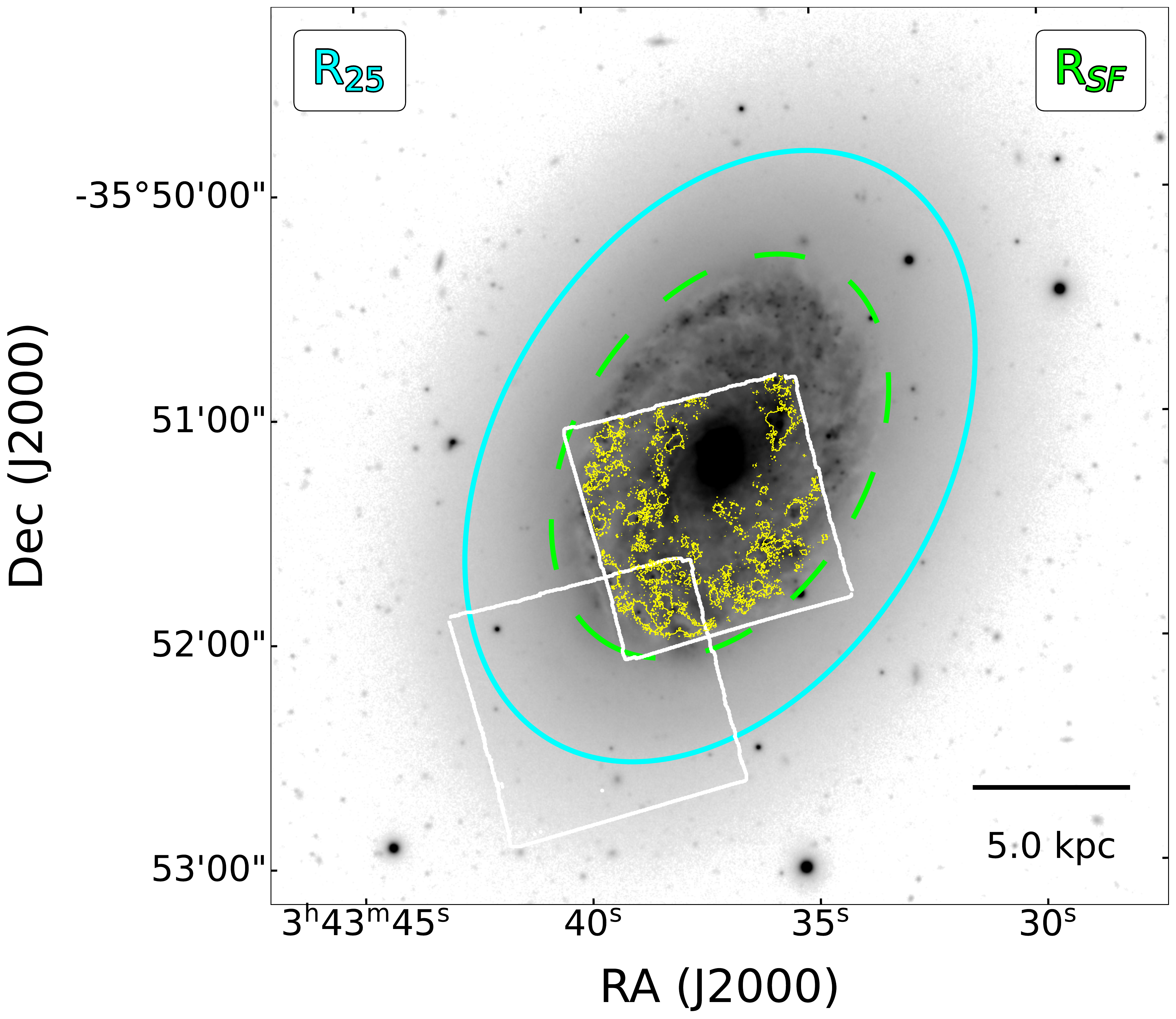} 
     \caption{Optical $g-$band image of NGC~1436 from the Fornax Deep Survey \citep{Reynier_2020arXiv200812633P} showing the morphological distinction between the inner star-forming regions characterised by flocculent spiral features and the quiescent and smoother outer disc. The footprint of the Fornax3D observations for NGC~1436 \citep{sarzi2018A&A...616A.121S} (white squares) and the map of the 
     H$\alpha$ emission detected in the MUSE data by \citealp[]{iodiceFDS2019} (yellow) are also shown. The latter measurements set the extent of the star-forming disc ($R_\mathrm{SF}~=~59$~\arcsec, dashed green line), while 
     the cyan isophotal radius measured at 25~mag~arcsec$^{-2}$ in B-band \citep{1991rc3..book.....D} gives an overall indication for the extent of NGC~1436.}
    \label{fig:opt_r25}
\end{figure}{}}

\newcommand{\placefigmom}{
\begin{figure*}
    \begin{center}
        \begin{subfigure}{0.47\textwidth}
            \includegraphics[width=\linewidth]{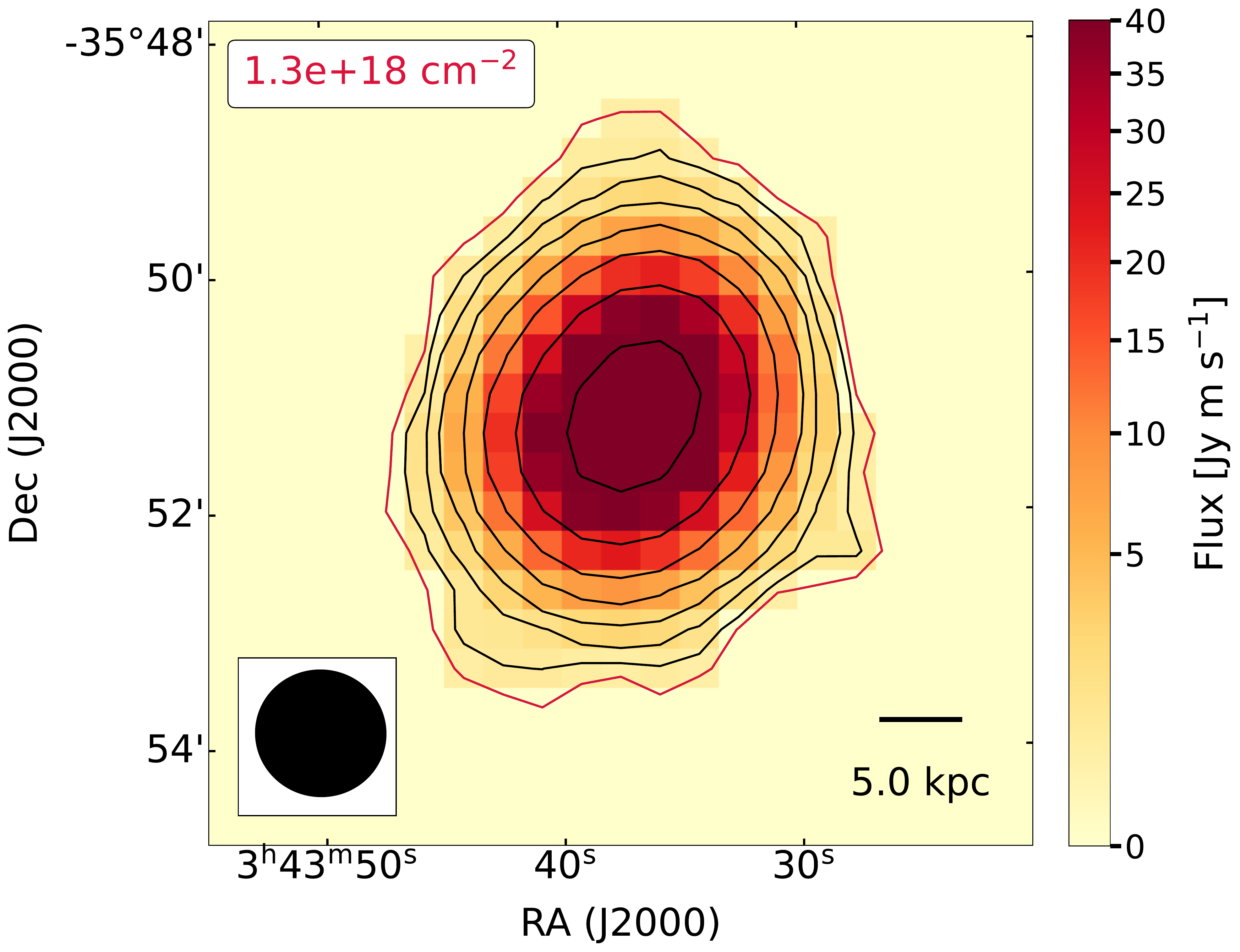}
        \end{subfigure}%
        \begin{subfigure}{0.48\textwidth}
             \includegraphics[width=\linewidth]{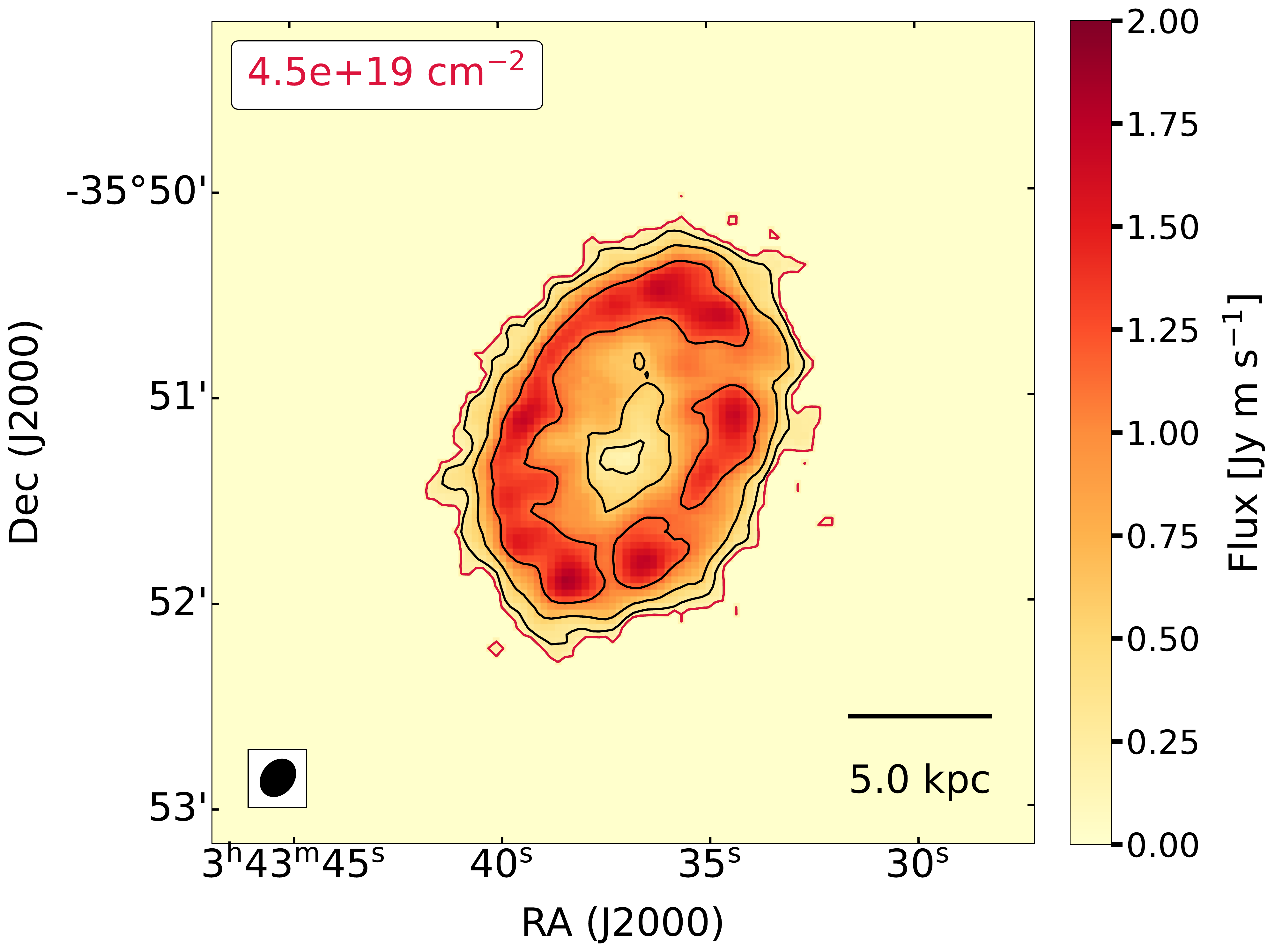}   
        \end{subfigure}\\ %
        \begin{subfigure}{0.40\textwidth}
            \includegraphics[width=\linewidth]{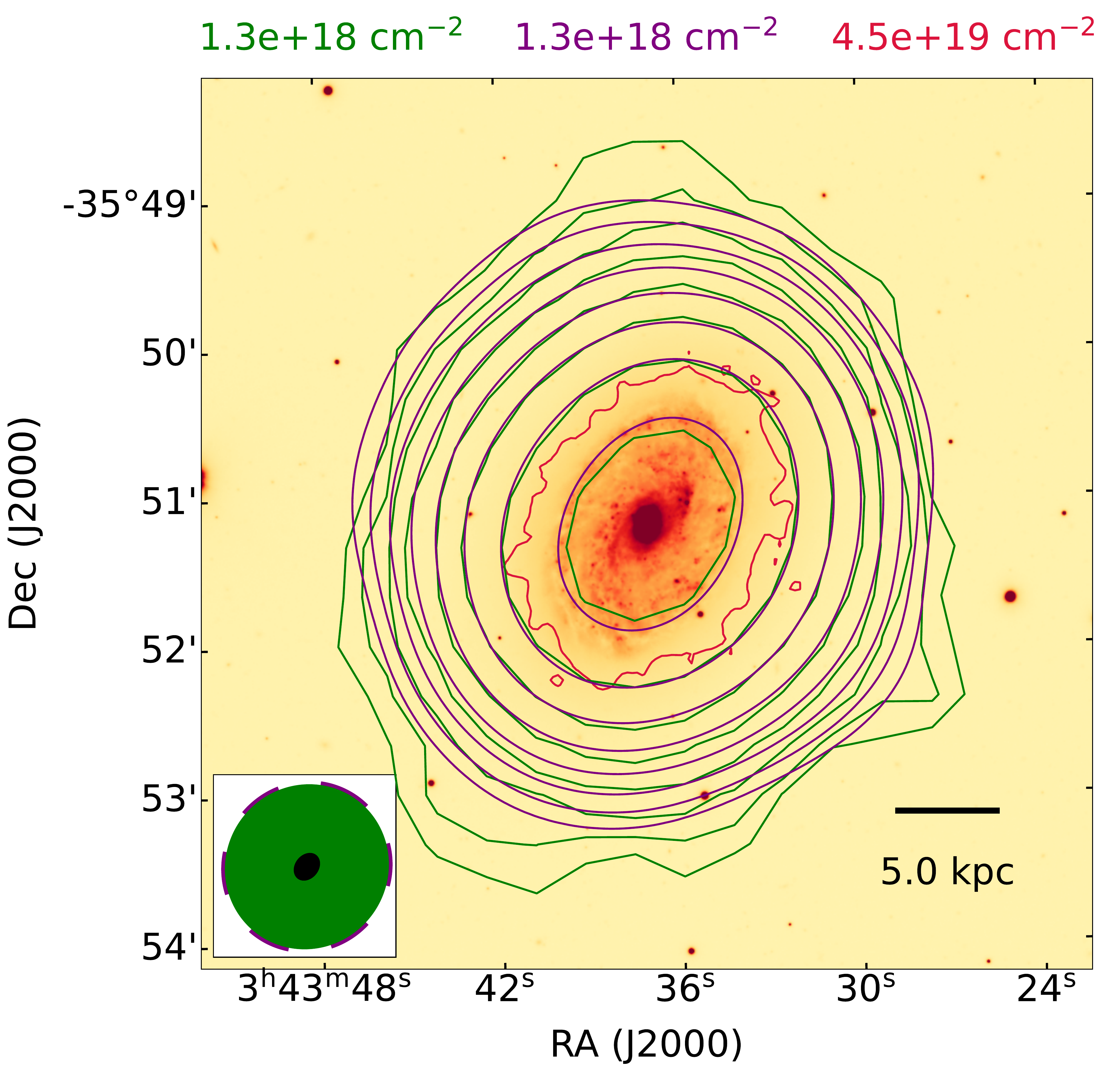}
        \end{subfigure}%
        \hspace{0.6cm}
        \begin{subfigure}{0.5\textwidth}
            \includegraphics[width=\linewidth]{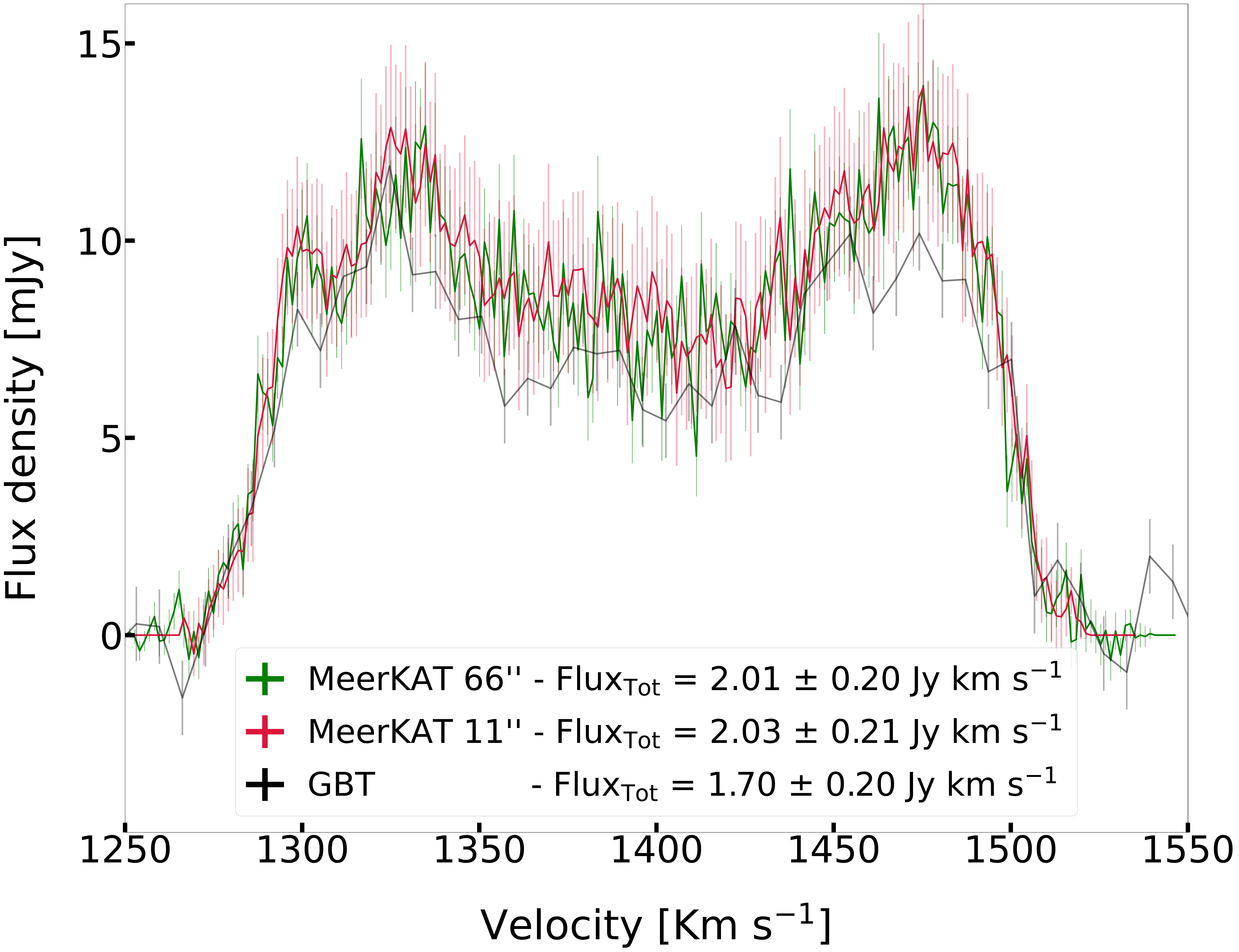}
        \end{subfigure}%
        \end{center}
        \caption{The top panels show the MeerKAT \hi{} image at a resolution of 66\arcsec (left) and 11\arcsec (right). In each panel the value on the top left gives the level of the lowest \hi{} contour (shown in red), which is calculated as the $3\sigma$ sensitivity over a 25\kms line width. The \hi{} contours increase by a factor of 2 at each step. The PSF and a 5~kpc scale bar are shown on the bottom-left and bottom-right corner, respectively. 
        The bottom left panel shows with green, red and purple contours the (\emph{1}) 66\arcsec \hi{} image (\emph{2}) lowest reliable contour of the 11\arcsec \hi{} image (\emph{3}) 11\arcsec image convolved to a 66\arcsec resolution. With the same colors we show the 3$\sigma$ column density sensitivity of the respective \hi{} image (top of the image) and their PSFs (bottom-left). The $g$-band optical image comes from the Fornax Deep Survey \citep{Reynier_2020arXiv200812633P}. The bottom right panel shows spectra extracted from the 11\arcsec (red) and 66\arcsec (green) cubes, and from single-dish data (GBT, \citealp{courtoisGBT} - black), respectively. The legend shows the total fluxes obtained from these data. The uncertainties on the total MeerKAT fluxes are calculated by summing in quadrature the statistical uncertainties -- derived from the local RMS and the number of independent pixels detected in each channel -- and the 10$\%$ MeerKAT flux uncertainty measured in \citet{2023arXiv230211895S}. The spectral axis shows velocity in the optical convention.}
        \label{fig:mom0s}
\end{figure*}}

\newcommand{\placefigPV}{\begin{figure*}
    \begin{center}

        \begin{subfigure}{0.45\linewidth}
            \includegraphics[width=\linewidth]{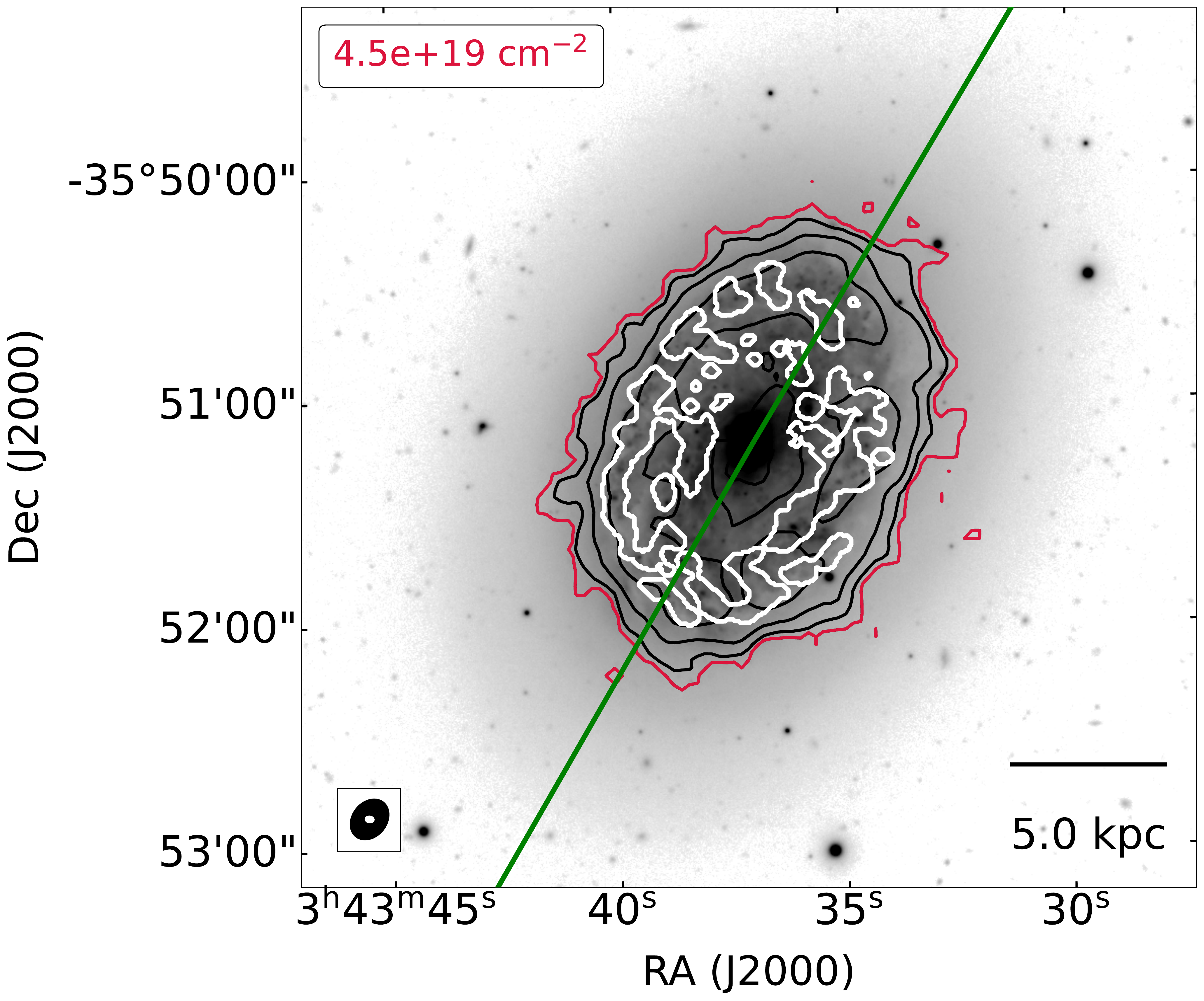}
        \end{subfigure}\hfill%
        \begin{subfigure}{0.54\linewidth}
             \includegraphics[width=\linewidth]{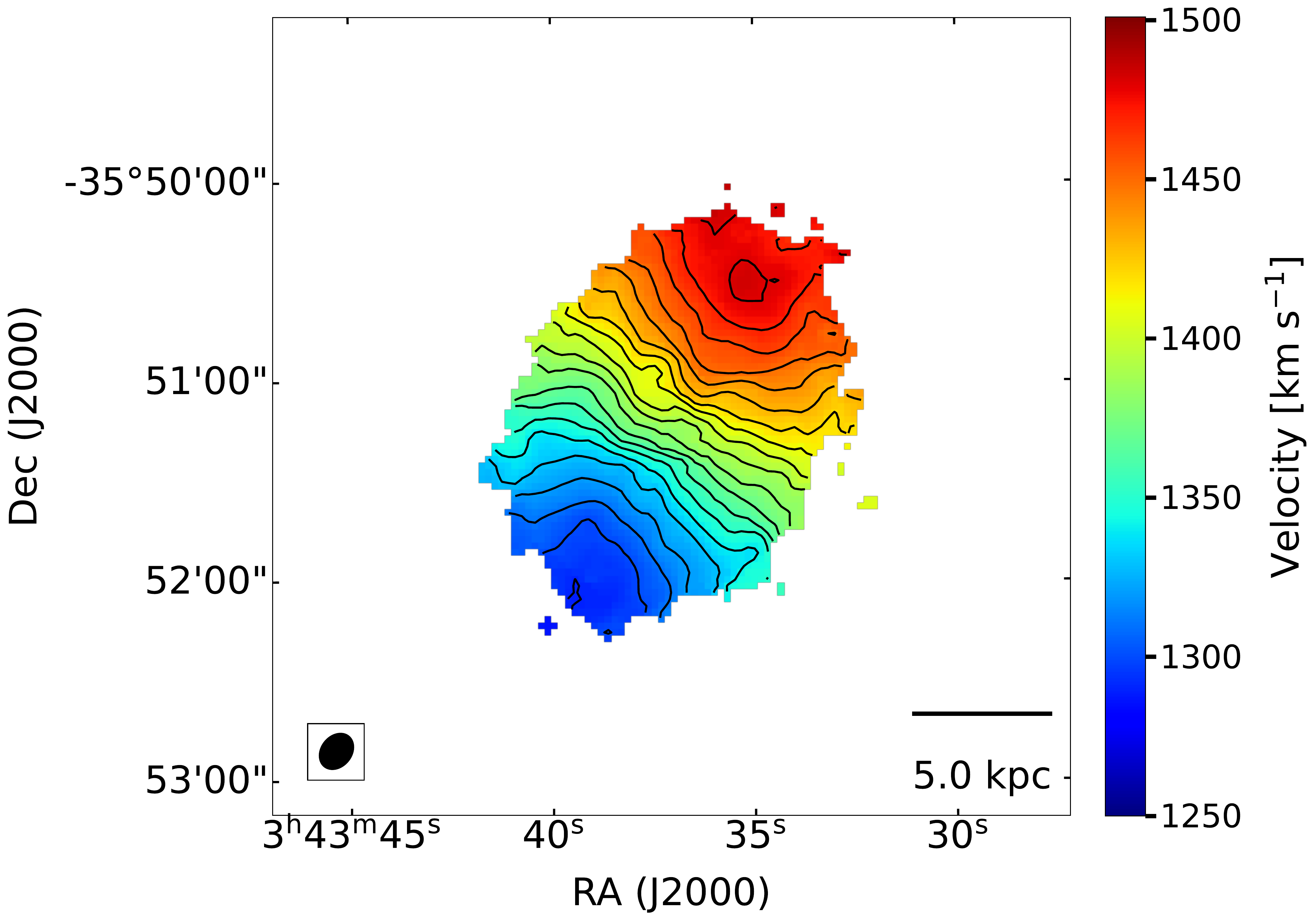}
             \end{subfigure}\\
        \begin{subfigure}{0.445\linewidth}
            \includegraphics[width=\linewidth]{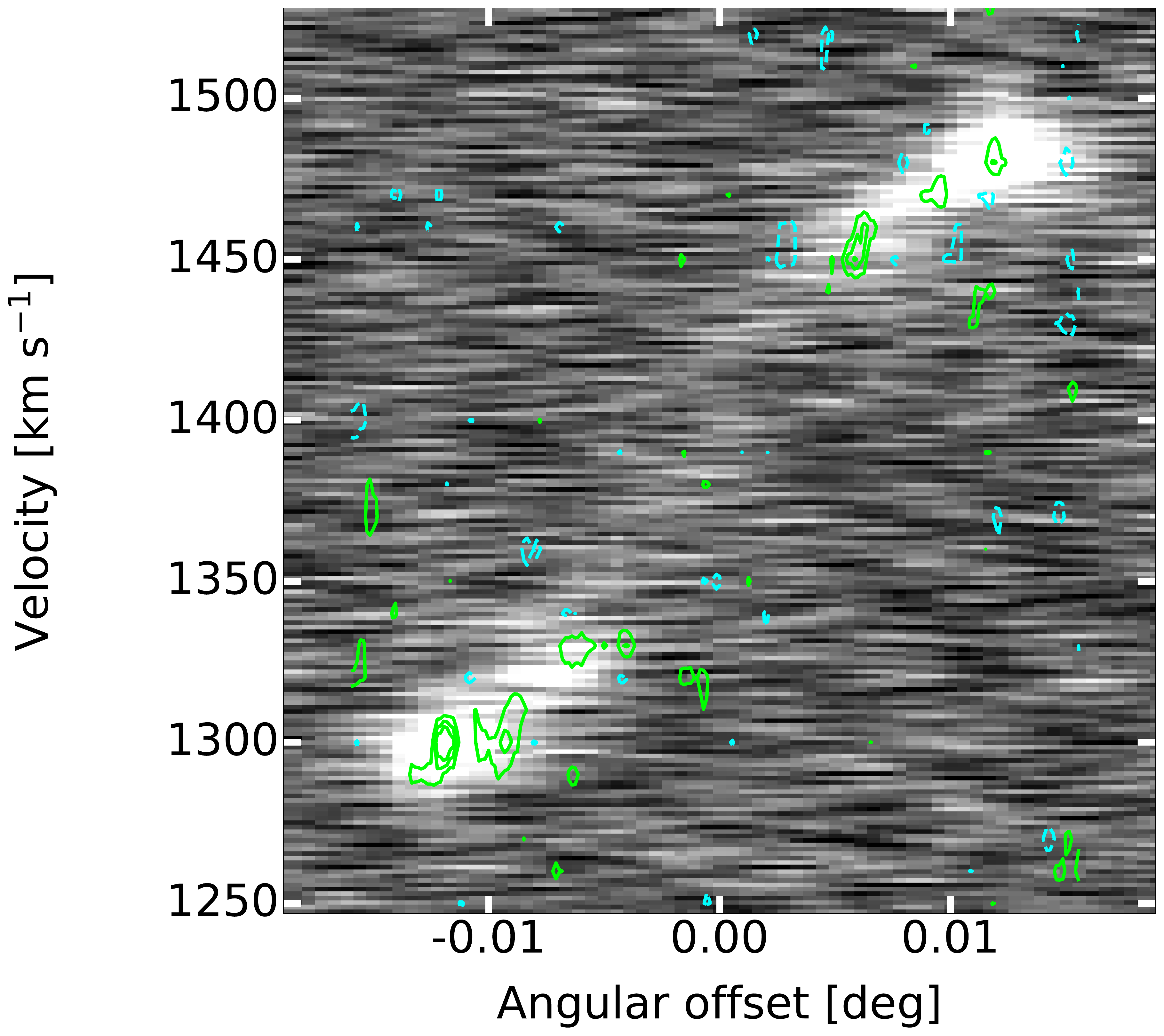}
        \end{subfigure}\hfill
        \begin{subfigure}{0.50\linewidth}
             \includegraphics[width=\linewidth]{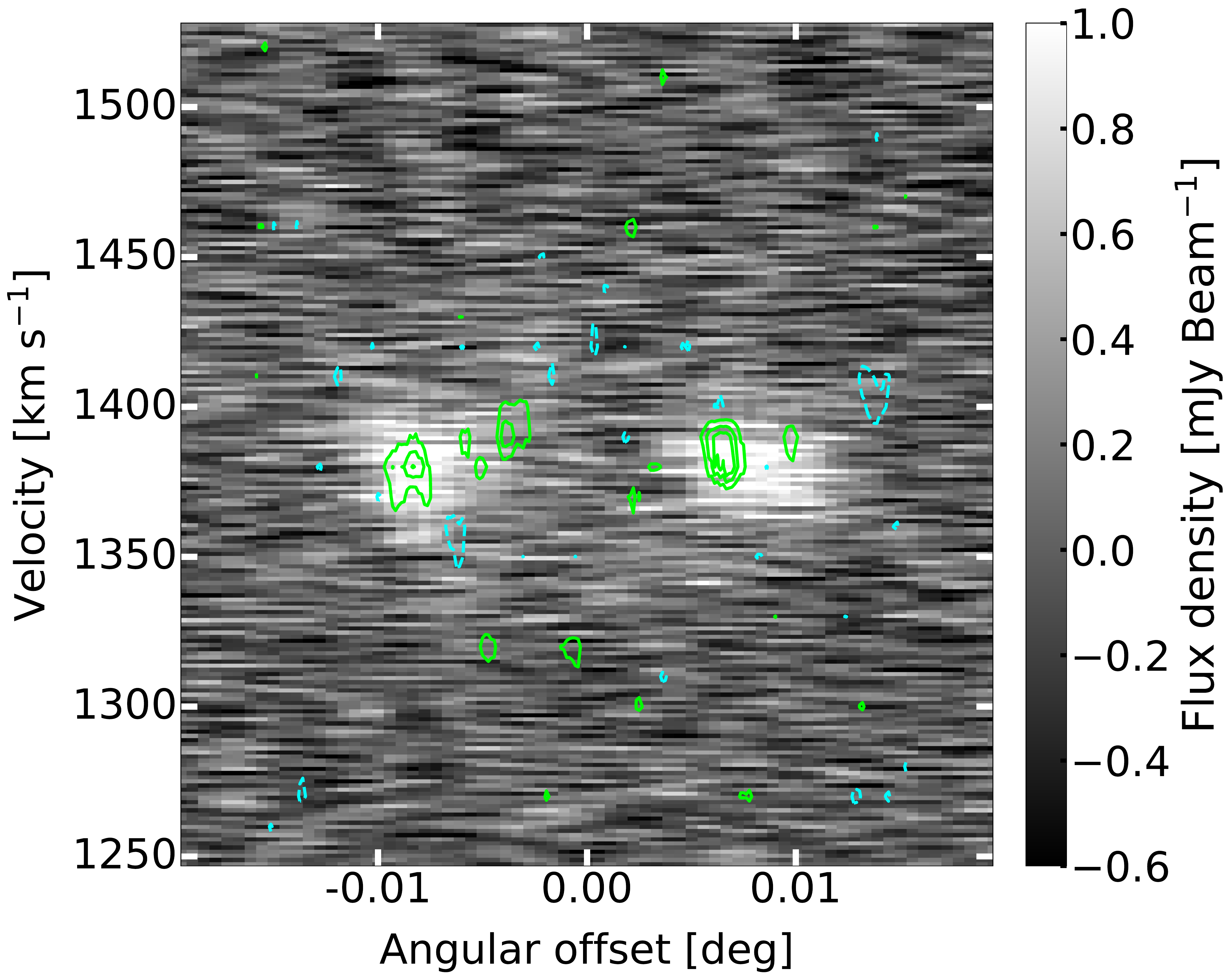}        
             \end{subfigure}
        \end{center}
        \caption{
        Top-left: comparison between the \hi{} and \htwo{} gas distribution from the MeerKAT (red and black contours as in the top-right panel of Fig.~\ref{fig:mom0s}) and ALMA (white contours) data, with corresponding beam sizes shown in the lower-left corner of this panel (in black and white, respectively).  
        Bottom panels: comparison of the PV-diagram of \hi{} and \htwo{} along the major axis (green line on the top left panel) and minor axis. The background image shows the \hi{} while the green contours show the \htwo{} emission at $2.5\sigma$, increasing in steps of $2$. Cyan contours show the negative signal in the \htwo{} ALMA cube, whose absolute level is the same as that of the first positive green contour.} 
        \label{fig:HIH2comp}
\end{figure*}}

\newcommand{\placefigTir}{\begin{figure}
    \centering
    \includegraphics[width=0.99\linewidth]{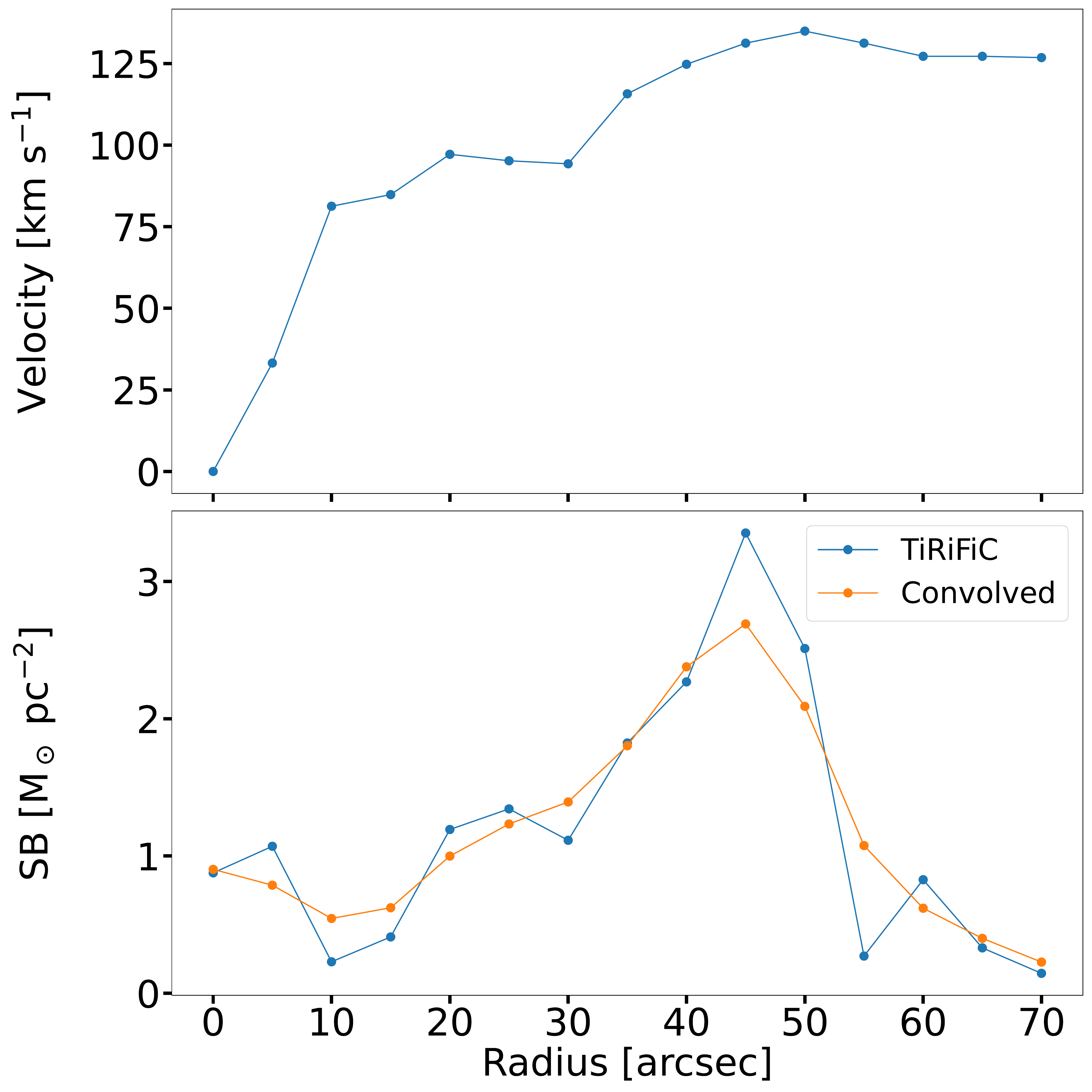}
     \caption{Results from tilted-ring analysis of the MeerKAT \hi{} data, showing the TiRiFiC \hi{} rotation curve (top panel) and \hi{} surface-brightness profile (bottom panel). The latter shows clearly the abrupt transition towards the \hi{} deficient and quiescent outer disc beyond a radius of 55\arcsec.}
    \label{fig:SBR}
\end{figure}{}}

\newcommand{\placefigMUSEHalphaapp}{\begin{figure}
    \centering{\includegraphics[width=\linewidth]{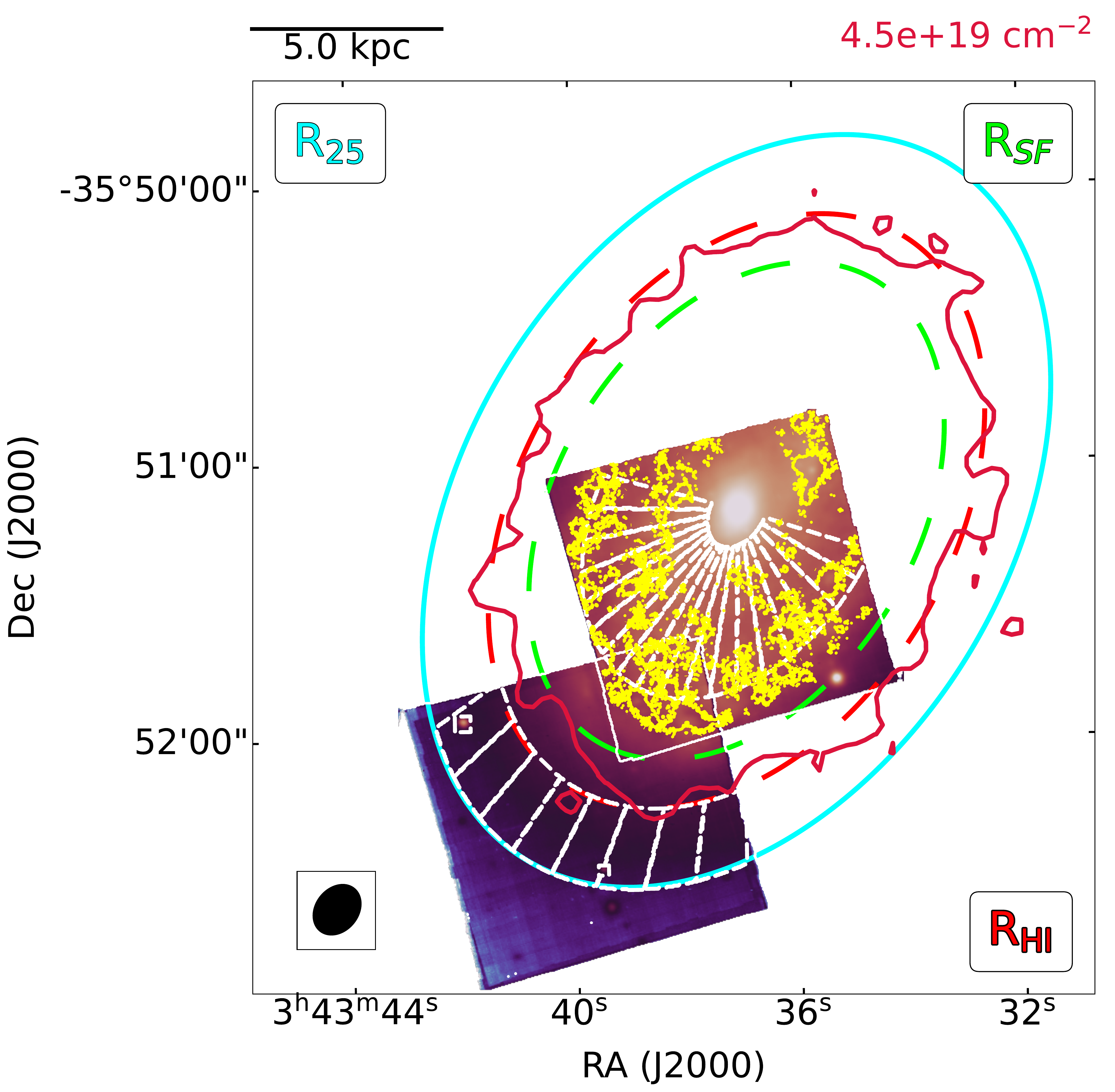}}
     \caption{Comparison between the reconstructed MUSE image, MUSE H$\alpha$ emission and the MeerKAT \hi{} data for NGC~1436 informing the location of the MUSE aperture spectra used in our star-formation history analysis. 
     As in Fig.~\ref{fig:opt_r25} the cyan and green ellipses indicated, respectively, the isophotal level at 25~mag/arcsec$^2$ in B-band and the limit of the star-forming disc, where the H$\alpha$ emission is traced by the yellow contours.
     The extent of the gas reservoir in NGC~1436 is traced here by the red contours for a limiting \hi{} column density of ${4.5\times10^{19}\rm cm^{-2}}$, which are best fitted by the shown red ellipse setting the radius of the \hi{} disc at $R_{\mathrm{HI}}~=~70~$\arcsec.    
     Finally, the white dashed lines show the location of the annular sectors used to extract the aperture spectra in both the inner and outer disc for our SFH analysis, where the former avoid the bulge regions (inside 9\arcsec) and extend to the limit of the MUSE central pointing whereas the latter start at the edge of the \hi{} disc and reach out to the 25~mag/arcsec$^2$ B-band isophote.
    \label{fig:SFH_regions}     
     }
\end{figure}{}}

\newcommand{\placefigspe}{\begin{figure*}
    \begin{center}
        \begin{subfigure}{0.49\linewidth}
            \includegraphics[width=\linewidth]{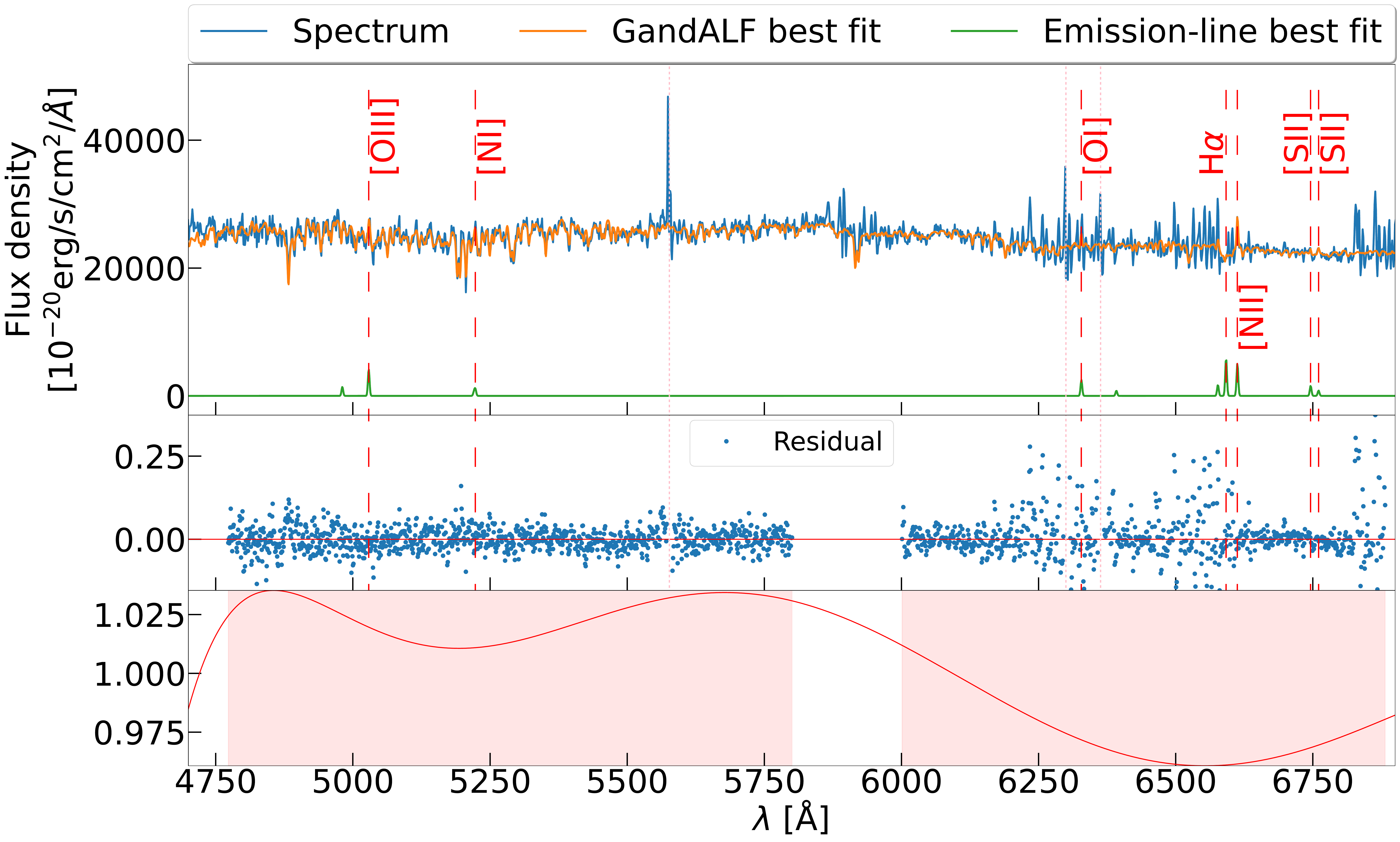}
        \end{subfigure}\hfill%
        \begin{subfigure}{0.49\linewidth}
             \includegraphics[width=\linewidth]{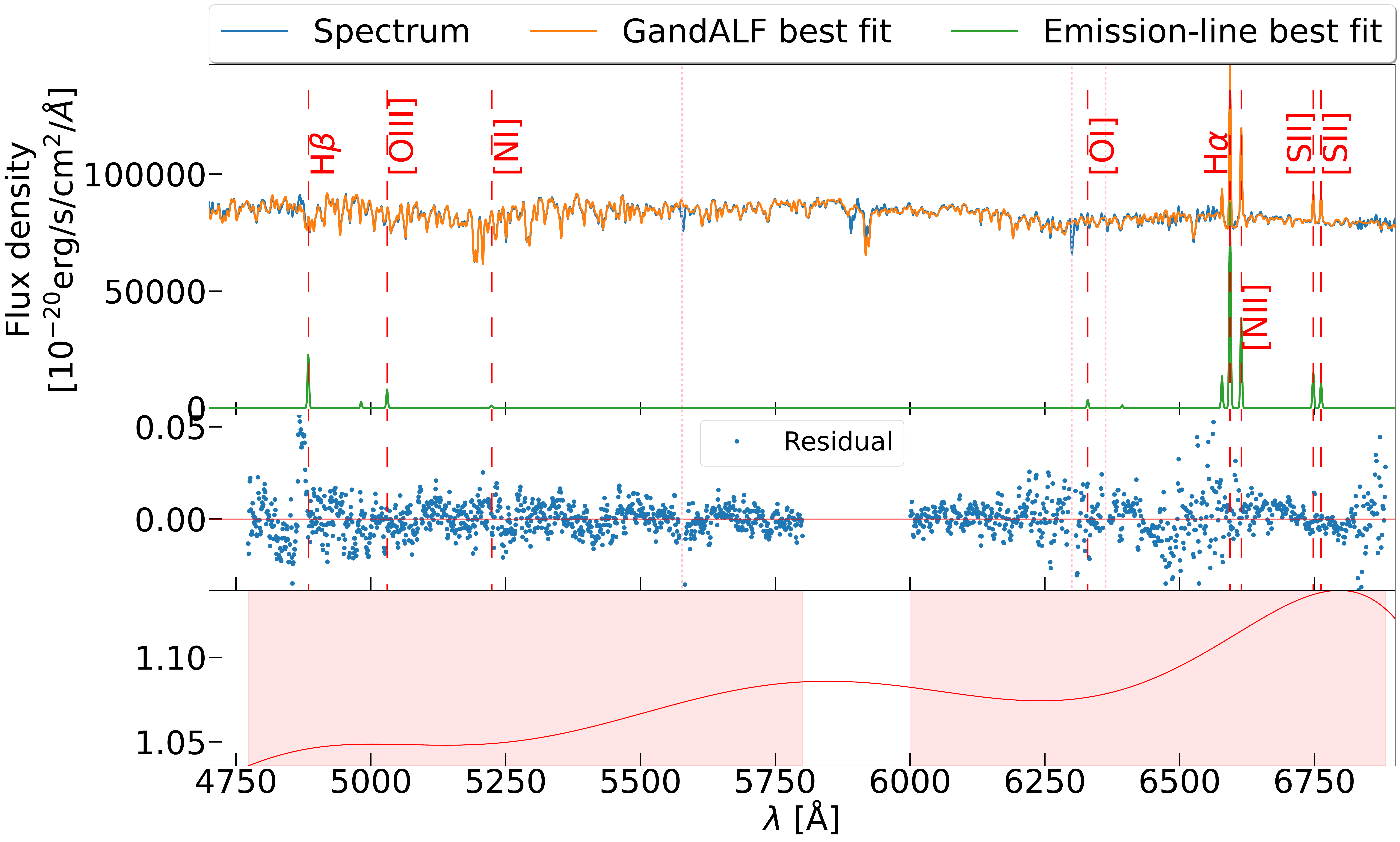}
        \end{subfigure}
    \end{center}
\caption{
Examples of GandALF (orange) to the MUSE spectra (blue) extracted in the outer (left) and inner disc (right). 
The green line shows the emission line best fit. The middle panel shows the residuals of the fit. The gap within the residuals indicates that we excluded that wavelength range from the GandALF fit. The vertical dashed red lines show the emission lines we fitted. In both cases shown, we used a 7-th order polynomial, whose shapes are shown in the bottom panels, respectively.}\label{fig:MUSEdatapoly}
\end{figure*}}

\newcommand{\placefigsinglesfh}{\begin{figure*}
    \begin{center}
        \begin{subfigure}{0.50\linewidth}
            \includegraphics[width=\linewidth]{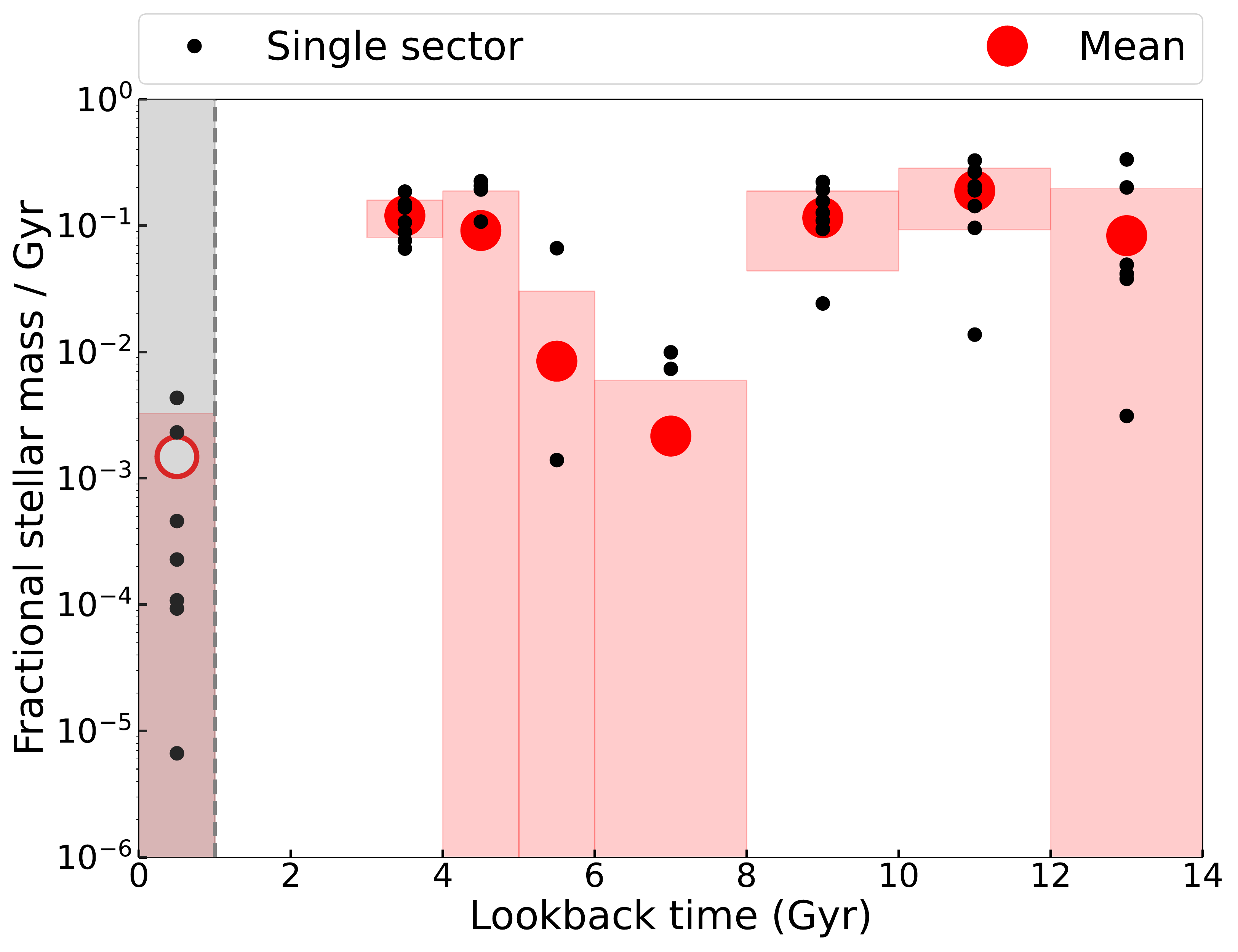}
        \end{subfigure}
        \begin{subfigure}{0.50\linewidth}
             \includegraphics[width=\linewidth]{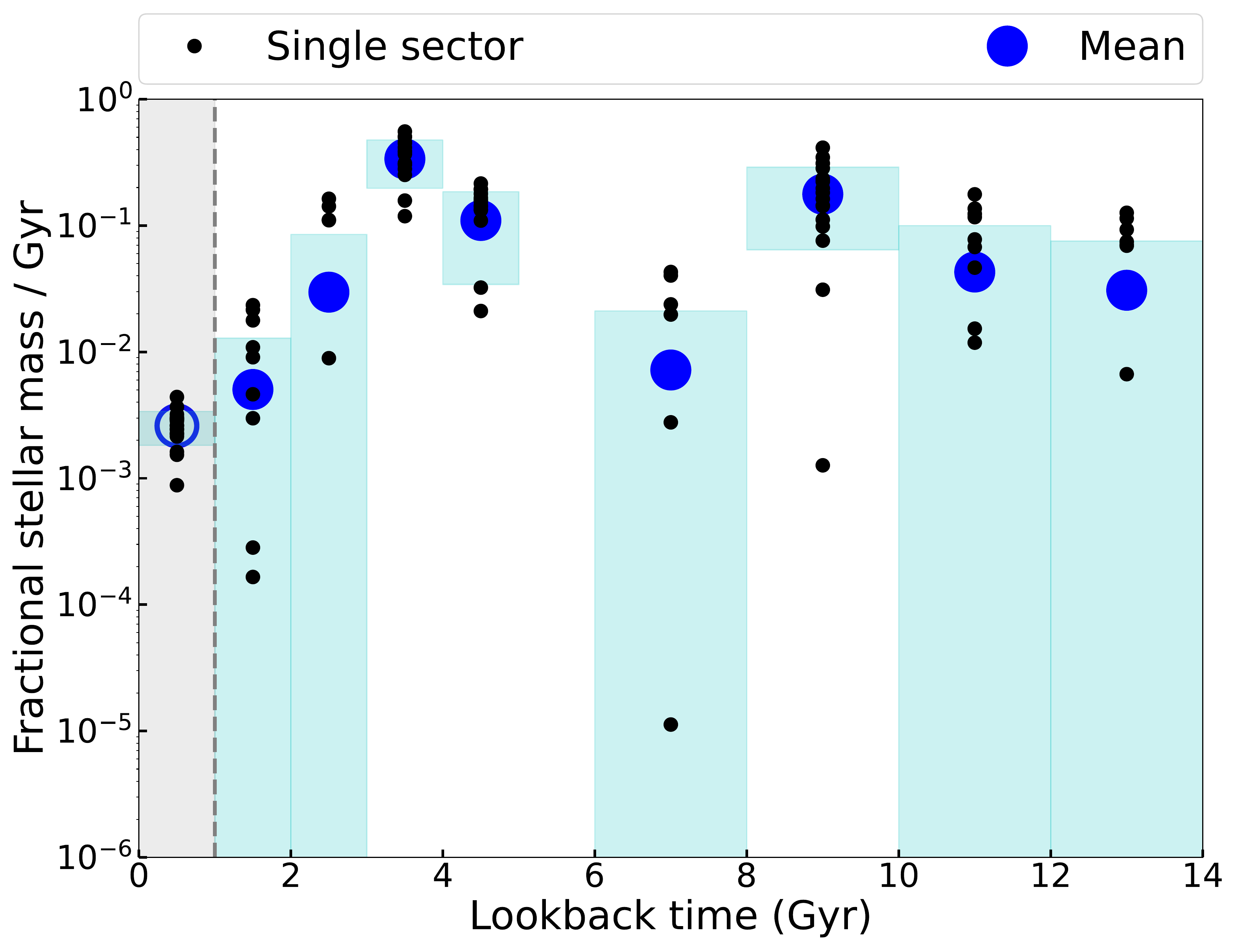}  
             \end{subfigure}
        \end{center}
        \caption{
        Sample normalised SFH of the quiescent (left) and the star-forming disc (right) of NGC~1436 obtained combining the results from the 8 and 18 sectors shown in Fig.~\ref{fig:SFH_regions} and while useing a 7-th order multiplicative polynomial correction for the continuum shape as illustrated in the examples of Fig.~\ref{fig:MUSEdatapoly}.
        In both panels the black dots represent the average fractional stellar mass formed within each age bin in each of our disc sectors, whereas the larger filled circles and shaded regions indicate the overall averages and standard deviations for the fractional SFRs, calculated including also zero fractional SFR values from individual sectors. The empty circles and gray bands highlight the unreliable measurements of the fractional SFR at younger ages (see text).}
        \label{fig:SFHs_oneexp}
\end{figure*}}

\newcommand{\placefigfinalsfh}{\begin{figure*}
    \begin{center}
        \begin{subfigure}{0.50\linewidth}
            \includegraphics[width=\linewidth]{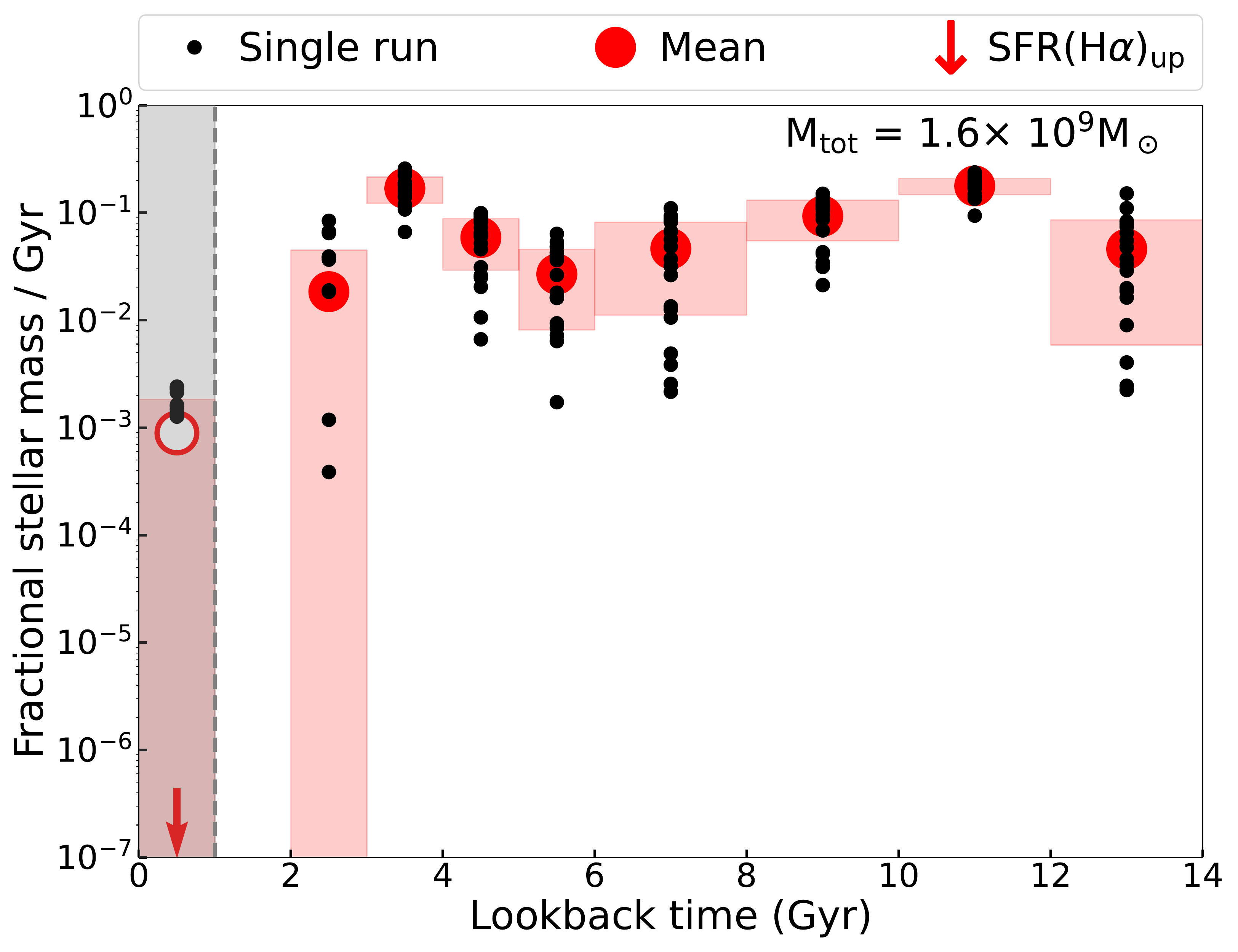}
        \end{subfigure}
        \begin{subfigure}{0.50\linewidth}
             \includegraphics[width=\linewidth]{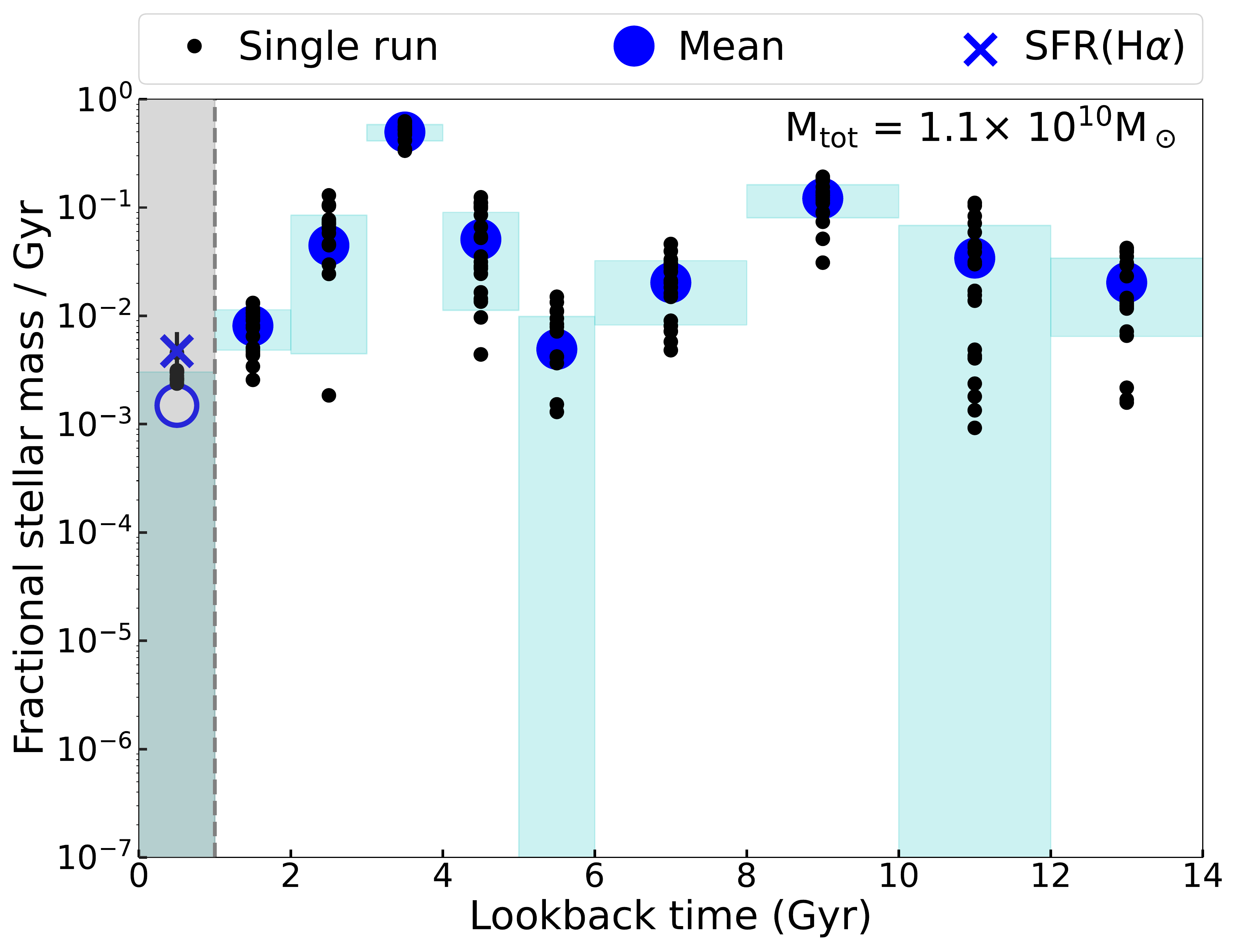}        
             \end{subfigure}
        \end{center}
          \caption{
          Final normalised SFH for the quiescent (left) and the star-forming disc (right) obtained by combining the GandALF results of 20 different fit runs that \emph{(i)} varied the order of the multiplicative polynomial correction between 6 and 15 and \emph{(ii)} considered either the entire MILES SSP library or only SSP models older than 1 Gyr.
          In both panels the black dots represent the average fractional SFR obtained from the individual fit runs (the same as the large filled circles in the example shown Fig.~\ref{fig:SFHs_oneexp}.
          The larger filled circles and shaded regions represent the average and standard deviation of the average fractional SFR of the single runs, including those equal to 0. 
          For the earliest age bin, where our SFH reconstruction is unreliable (open circles and shaded grey band), we uses the MUSE H$\alpha$ maps to measure or place an upper limit on the present-day star-formation rate across our inner or outer disc sectors, respectively, converting these in estimate (blue cross, right panel) or upper-limits (red downward arrow, left panel) on the fractional SFR of the first 1 Gyr by assuming such an activity remained constant over this period of time.
          On the top right of each panel we report the total M$_\star$ subtended by our 8 outer and 18 inner disc sectors, which cover an area of 12.5~kpc$^2$ and 18~kpc$^2$, respectively. }
        \label{fig:SFHs_allexp}
\end{figure*}}

\newcommand{\placeTNGbothsamecap}{\begin{figure*}
    \begin{center}
        \begin{subfigure}{0.465\linewidth}
            \includegraphics[width=\linewidth]{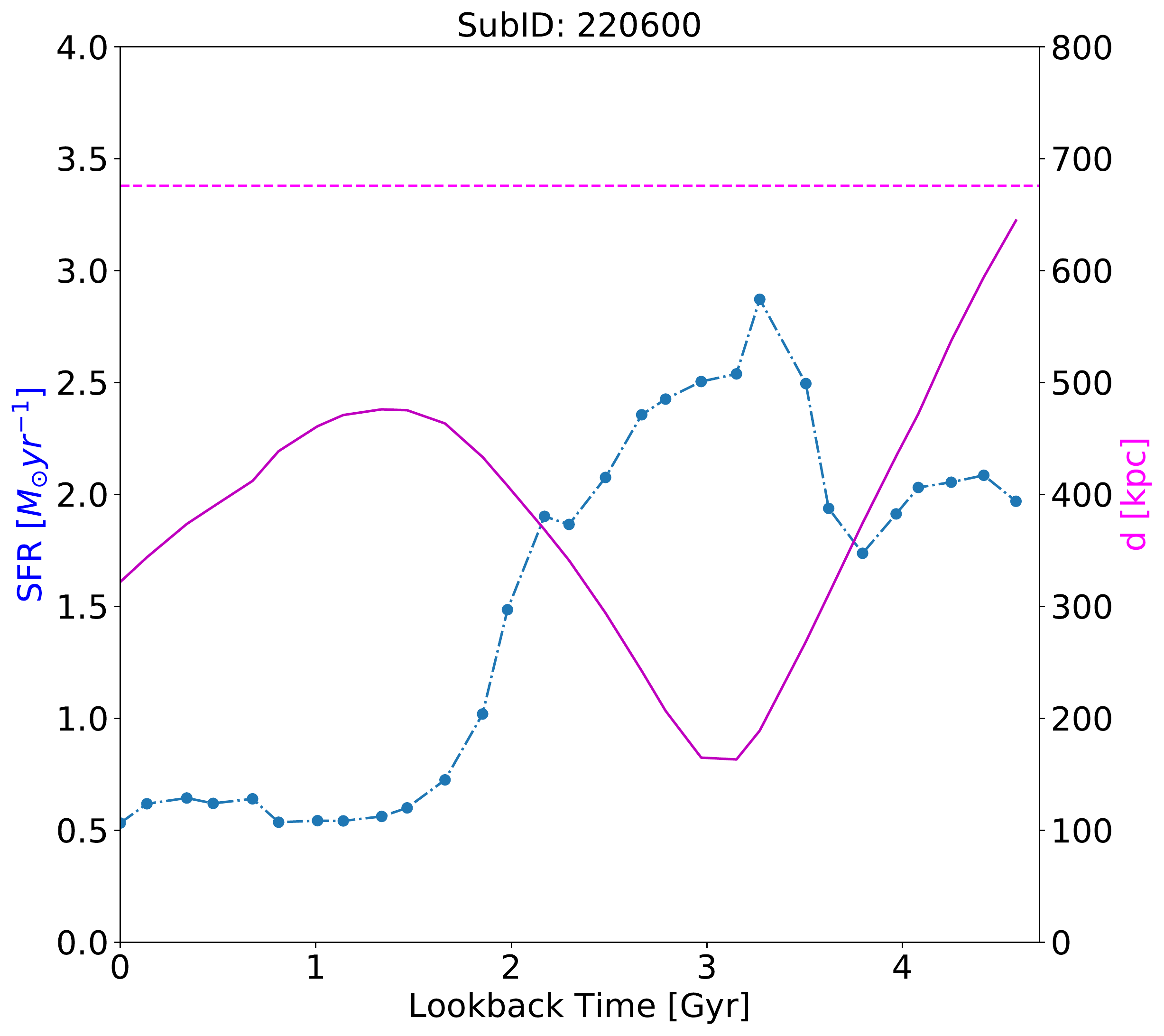}
        \end{subfigure}
        \begin{subfigure}{0.50\linewidth}
             \includegraphics[width=\linewidth]{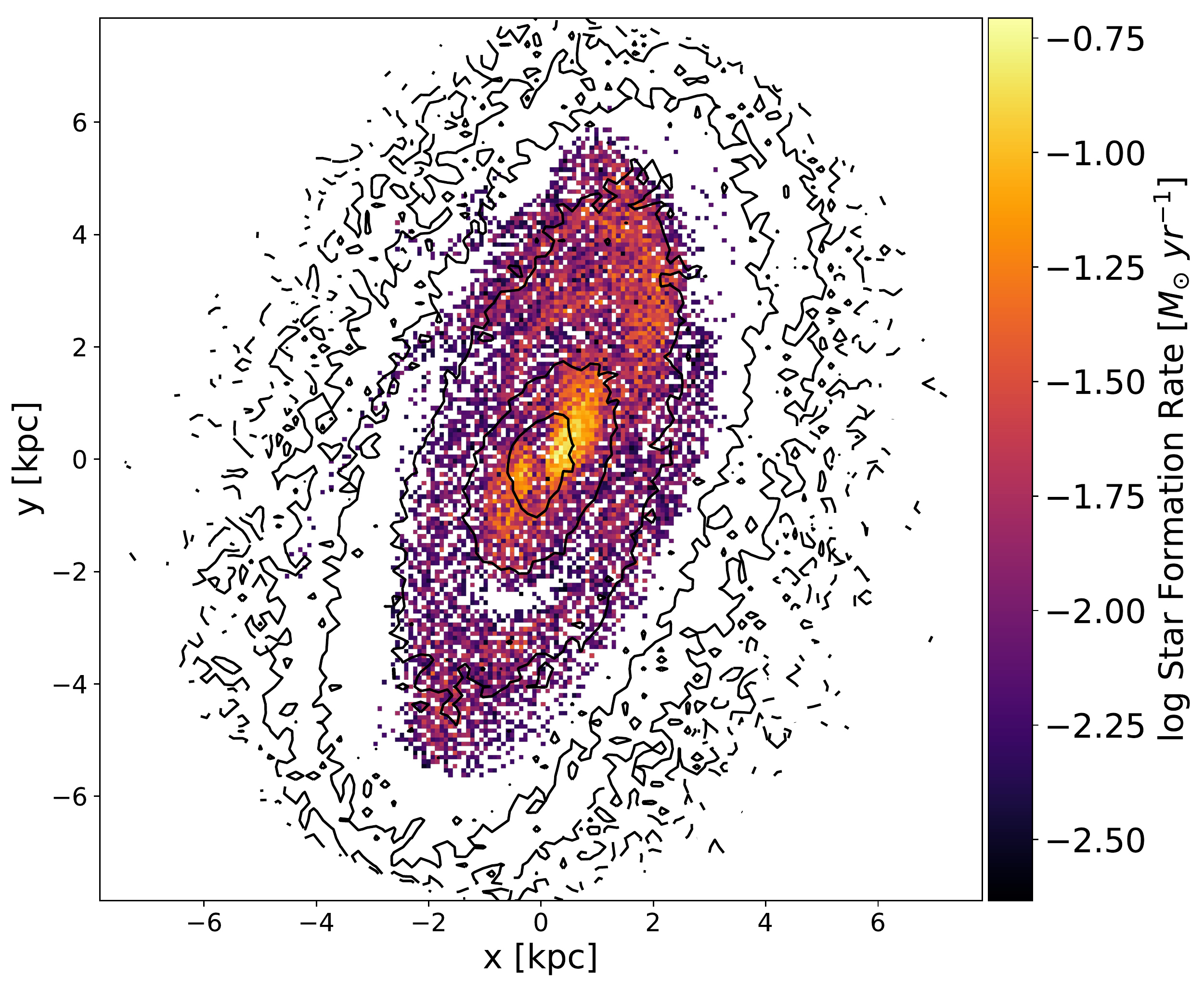}  
             \end{subfigure}
        \end{center}
        \caption{
        The left panel shows an example from the TNG50 simulation run \citep{Pillepich19} of a disk galaxy that recently fell in a Fornax-like cluster environment in this simulation \citep{PabloTNG50} identified through the SUBFIND algorithm \citep{2001MNRAS.328..726S}. As inferred from our SFH analysis for NGC~1436, this simulated object experienced a burst of star formation upon its first pericentre passage, followed by a steady drop in star-formation activity. The purple line traces the radial distance from the cluster center as a function of time, whereas the dot-dashed blue line follows the evolution of star-formation activity. The horizontal dashed purple line indicate the virial radius of this particular Fornax-like TNG50 cluster. The right panel shows the SFR map of the target TNG50 disc galaxy at \emph{z}=0. Black contours show the surface mass density which highlight that, similarly to NGC~1436, the star formation is occurring just within the inner part of the disc.}
        \label{fig:sfr_tng}
\end{figure*}}

\title[NGC 1436: the making of a lenticular galaxy in the Fornax cluster]{NGC 1436: the making of a lenticular galaxy in the Fornax cluster}

\author[A. Loni et al.]{Alessandro Loni,$^{1,2}$\thanks{E-mail: alessandro.loni@armagh.ac.uk (AL)}
Paolo~Serra,$^{2}$
Marc~Sarzi,$^{1}$ 
Gyula~I.~G.~Józsa,$^{3,4}$
Pablo M. ~Gal\'{a}n$-$de~Anta,$^{1,5}$
Nikki~Zabel,$^{6}$
\newauthor
Dane~Kleiner,$^{7,2}$
Filippo~M.~Maccagni,$^{7,2}$
Daniel~Molnár,$^{2}$
Mpati~Ramatsoku,$^{4,2}$
Francesca~Loi,$^{2}$
\newauthor
Enrico~M.~Corsini,$^{8,9}$
D.~J.~Pisano,$^{6}$
Peter~Kamphuis,$^{10}$
Timothy~A.~Davis,$^{11}$
W.~J.~G.~de~Blok,$^{7,12,13}$
\newauthor
Ralf~J.~Dettmar,$^{10}$
Jesus~Falcon-Barroso,$^{14,15}$
Enrichetta~Iodice,$^{16}$
Maritza~A.~Lara-L\'{o}pez,$^{17}$
\newauthor
S.~Ilani~Loubser,$^{18}$
Kana~Morokuma-Matsui,$^{19}$
Reynier~Peletier,$^{12}$
Francesca~Pinna,$^{20}$
Adriano~Poci,$^{21}$
\newauthor
Matthew~W.~L.~Smith,$^{11}$
Scott~C.~Trager,$^{12}$
and
Glenn~van de Ven$^{22}$
\\
$^{1}$Armagh Observatory and Planetarium, College Hill, Armagh BT61 9DG, UK\\
$^{2}$INAF $-$ Osservatorio Astronomico di Cagliari, Via della Scienza 5, 09047, Selargius, CA, Italy\\
$^{3}$Max-Plank-Institut f\"ur Radioastronomie, Auf dem Hügel 69, 53121 Bonn, Germany\\
$^{4}$Department of Physics and Electronics, Rhodes University, PO Box 94, Makhanda 6140, South Africa\\
$^{5}$Astrophysics Research Centre, School of Mathematics and Physics, Queen’s University Belfast, Belfast BT7 INN, UK\\
$^{6}$Department of Astronomy, University of Cape Town, Private Bag X3, Rondebosch 7701, South Africa\\
$^{7}$Netherlands Institute for Radio Astronomy (ASTRON), Oude Hoogeveensedijk 4, 7991 PD Dwingeloo, the Netherlands \\
$^{8}$Dipartimento di Fisica e Astronomia ``G. Galilei'', Universit\`a di Padova, vicolo dell'Osservatorio 3, I-35122 Padova, Italy\\
$^{9}$INAF - Osservatorio Astronomico di Padova, vicolo dell'Osservatorio 2, I-35122 Padova, Italy\\
$^{10}$Ruhr University Bochum, Faculty of Physics and Astronomy, Astronomical Institute, 44780 Bochum, Germany\\
$^{11}$School of Physics and Astronomy, Cardiff University, Queens Buildings The Parade, Cardiff, CF24 3AA, UK\\
$^{12}$Kapteyn Astronomical Institute, University of Groningen, PO Box 800, NL$-$9700 AV Groningen, the Netherlands\\
$^{13}$The Inter-University Institute for Data Intensive Astronomy, Department of Astronomy, University of Cape Town,\\~~~  Private Bag X3, Rondebosch, 7701, South Africa\\
$^{14}$Instituto de Astrof\'isica de Canarias, V\'ia L\'actea s/n, E-38205 La Laguna, Tenerife, Spain\\
$^{15}$Departamento de Astrof\'isica, Universidad de La Laguna, E-38200 La Laguna, Tenerife, Spain\\
$^{16}$INAF $-$ Astronomical Observatory of Capodimonte, Salita Moiariello 16, 80131, Naples, Italy\\
$^{17}$Departamento de F\'{i}sica de la Tierra y Astrofísica, Instituto de F\'{i}sica de Part\'{i}culas y del Cosmos, IPARCOS.\\~~~ Universidad Complutense de Madrid (UCM), 28040, Madrid, Spain\\
$^{18}$Centre for Space Research, North$-$West University, Potchefstroom 2520, South Africa\\
$^{19}$Center for Computational Sciences, University of Tsukuba, Ten-nodai, 1-1-1 Tsukuba, Ibaraki 305-8577, Japan\\
$^{20}$Max-Planck-Institut f\"{u}r Astronomie, K\"{o}nigstuhl 17, D-69117 Heidelberg, Germany\\
$^{21}$Centre for Extragalactic Astronomy, University of Durham, Stockton Road, Durham DH1 3LE, United Kingdom\\
$^{22}$Department of Astrophysics, University of Vienna, T\"urkenschanzstra{\ss}e 17, 1180 Vienna, Austria\\
}

\date{Accepted 2023 May 09. Received 2023 May 08; in original form 2023 March 02}

\pubyear{2023}

\begin{document}
\label{firstpage}
\pagerange{\pageref{firstpage}--\pageref{lastpage}}
\maketitle

\begin{abstract}
We study the evolutionary path of the Fornax cluster galaxy NGC~1436, which is known to be currently transitioning from a spiral into a lenticular morphology. This galaxy hosts an inner star-forming disc and an outer quiescent disc, and we analyse data from the MeerKAT Fornax Survey, ALMA, and the Fornax~3D survey to study the interstellar medium and the stellar populations of both disc components. 
Thanks to the combination of high resolution and sensitivity of the MeerKAT data, we find that the \hi{} is entirely confined within the inner star-forming disc, and that its kinematics is coincident with that of the \htwo{}. The cold gas disc is now well settled, which suggests that the galaxy has not been affected by any environmental interactions in the last $\sim1$~Gyr.
The star formation history derived from the Fornax~3D data shows that both the inner and outer disc experienced a burst of star formation $\sim5$ Gyr ago, followed by rapid quenching in the outer disc and by slow quenching in the inner disc, which continues forming stars to this day. We claim that NGC~1436 has begun to effectively interact with the cluster environment 5~Gyr ago, when a combination of gravitational and hydrodynamical interactions caused the temporary enhancement of the star-formation rate. Furthermore, due to the weaker gravitational binding \hi{} was stripped from the outer disc, causing its rapid quenching. At the same time, accretion of gas onto the inner disc stopped, causing slow quenching in this region.
\end{abstract}

\begin{keywords}
Galaxies: individual: NGC~1436 -- Galaxies: evolution -- Galaxies: ISM -- Galaxies: star formation -- Galaxies: cluster
\end{keywords}


\section{Introduction}
\label{sec:n1436intro}

Due to their discy structure and old stellar populations lenticular galaxies (S0s) share properties both with spirals and ellipticals \citep{1936rene.book.....H, 1976ApJ...206..883V}.
The distribution and fractional number of lenticular galaxies in the Universe at different epochs (lower at higher redshift and outside groups and clusters of galaxies) suggest that they form mainly (but not always, c.f. \citealp{2009ApJ...702.1502V}) from the evolution of blue star-forming spiral galaxies in dense environments \citep{dressler1980galaxy,dres_1997ApJ...490..577D,2000ApJ...542..673F,2005ApJ...623..721P}.
Although there are still many open questions about how this morphological transformation happens, it is clear that environmental interactions can play an important role (e.g.~\citealp{1983ApJ...264...24M,1996Natur.379..613M}).

A peculiar characteristic of the morphological evolution of galaxies in dense environments is that it proceeds outside-in. 
This transformation requires, in the first place, to stop cold gas accretion onto galaxies, which would otherwise supply material to form new stars. This process is commonly referred to as "starvation" \citep{1980ApJ...237..692L}.
Then, the cold gas, mainly made of neutral atomic hydrogen (\hi{}), may be actively removed from the outer disc, where the weaker gravitational potential is not able to hold it in place against the action of environmental interactions. Therefore, whereas the outskirts of galaxies, which are experiencing cold gas removal, are quenched quickly if no further cold gas is supplied, star formation can continue in the inner - and still \hi{} rich - part of these systems. The outside-in transformation of galaxies which live in dense environments is opposite to what happens to field and void galaxies, where accretion of gas from the intergalactic medium generally keeps fuelling star-formation in the outer disc (e.g. \citealp{2000MNRAS.312..497B,boselligavazzi,bosellifateofspirals2006ApJ...651..811B}). 

Through the study of the distribution and kinematics of the cold gas, one can gain information on which kind of interactions a galaxy might have experienced after its starvation has begun (if at all). On the one hand hydrodynamical interactions (e.g. ram pressure stripping \citealp{1972ApJ...176....1G,1999MNRAS.308..947A,2000Sci...288.1617Q}), which only affect the gas component, cause highly asymmetric perturbations in the cold gas distribution, producing \hi{} tails which trail along the orbit of the galaxy. On the other hand, the footprint of tidal interactions in the cold gas distribution and possibly in the stellar disc is more symmetric as one can see in some two tailed objects in \cite{1966apg..book.....A}, whose evolutionary scenarios are discussed and modelled by \cite{2012MNRAS.422.2444S}. Indeed, these interactions can cause two long streams of material in opposite directions (although they can have very different sizes).

The main difference between starvation and starvation plus active gas removal concerns different quenching time scales. Although in the literature there are several definitions of quenching timescale (see \citealp{Lucareview_2021PASA...38...35C} for a discussion), starvation implies longer time scales for a galaxy to halt its star formation (roughly 4~Gyr; \citealp{2014A&A...564A..66B}) with respect to the shorter time scales due to active gas removal (of the order of a few hundreds of Myr).
If the study of the cold gas distribution and kinematics can help us understand the cause of relatively recent gas removal, the analysis of the stellar populations and, thus, of the star formation history (SFH) of a galaxy is crucial to distinguish between slow (starvation-like) and fast (stripping-like) quenching.

One of the members of the nearby Fornax galaxy cluster (distance $\sim20$~Mpc; \citealp[]{2001ApJ...550..503J,2001ApJ...546..681T,blakeslee2009acs,Spriggs21}), NGC~1436 (FCC~290), is a perfect target to study "how" and "how quickly" the evolution of spirals into S0's can happen in dense environments. This galaxy is located $\sim0.5\times$R$_{\rm vir}$~=~350~kpc
south-east from the center of the cluster, and it has a systemic velocity of 1392\kms similar to the recessional velocity of the Fornax cluster (1442\kms, \citealp{Maddox2019}). Neither the 2D distribution of the Fornax galaxies nor their position in the projected phase space diagram \citep{IodiceF3D} show any obvious companion for NGC~1436. The closest galaxy with a known redshift is FCC~298, which is at $17$\arcmin $\sim100$~kpc,~ and $\sim$230\kms away from NGC~1436. Deep optical observations from the Fornax Deep Survey \citep{Reynier_2020arXiv200812633P} suggest that NGC~1436 is in the transitional stage to become a S0 \citep{iodiceFDS2019,FDS_RAJ}. 
In Fig.~\ref{fig:opt_r25} we see that the disc of NGC~1436 shows two distinct parts: a clumpy inner disc with a tightly wound spiral structure, abundant dust with numerous star-forming regions, and an outer disc which appears featureless (no dust, spiral arms, or star forming regions). \cite{FDS_RAJ} measured a break in the surface brightness slope at their boundary.

Data from the Fornax 3D survey \citep{sarzi2018A&A...616A.121S} obtained with the Multi-Unit Spectroscopic Explorer (MUSE) on the ESO Very Large Telescope also show how this galaxy is still forming stars in the inner disc (see the H$\alpha$ emission in Fig.~\ref{fig:opt_r25}) while its outskirts appear to be already quenched \citep{iodiceFDS2019}. This is further supported by the  Atacama Large Millimeter/submillimeter Array (ALMA) measurement of \cite{zabel}, which reveals the presence of molecular gas only within the inner disc. \cite{zabel2021MNRAS.502.4723Z} found that the radial profile of the \htwoVERO-to-dust ratio in NGC~1436 drops suddenly at a radius of 3.5~kpc, suggesting that environmental interactions were more efficient in removing dust rather than \htwoVERO.

\placefigopt

In this multi-wavelength picture of NGC~1436, high resolution \hi{} data were still missing. Indeed, the \hi{} resolution and sensitivity of the recent interferometric ATCA \hi{} survey of Fornax ($\sim1$\arcmin~and $\sim10^7$~M$_\odot$, respectively; \citealp[]{loni21}) are insufficient to \emph{(i)} establish whether the galaxy has recently interacted with the cluster environment \emph{(ii)} recover the same total \hi{} flux of a single dish (the ATCA \hi{} flux is 2.8$\times$ lower than that measured with the Green Bank Telescope -- by \citealp{courtoisGBT}).

\placefigmom

In this paper we present \hi{} data of NGC~1436 from the MeerKAT Fornax Survey \citep{2023arXiv230211895S}, whose resolution and sensitivity are at least a factor of 10 better than those of the previous ATCA \hi{} data.
We then compare the \hi{} distribution and kinematics with those of \htwo{} \citep{zabel}. This will establish whether the cold gas disc is well settled (if they are both distributed regularly and co-rotate), and thus whether NGC~1436 is currently in a state of slow evolution (no sign of recent interactions such as asymmetries). Finally, using MUSE data from the Fornax 3D survey \citep{sarzi2018A&A...616A.121S}, we measure and compare the SFH of the inner star-forming disc with that of the outer quiescent disc. By performing this analysis, we aim to understand in which epoch environmental interactions might have become relevant for this galaxy, and what consequences these had for its evolution.

The paper is organized as follows: in Sect~\ref{sec:observations} we present the MeerKAT, ALMA and MUSE data. In Sect.~\ref{sec:n1436gas} we analyse the distribution and kinematics of \hi{} and \htwo{}. In Sect.~\ref{sec:n1436sfhresult} we describe the MUSE data analysis and the measurement of the SFH of the quiescent and star-forming disc. In Sect.~\ref{sec:n1436discussion} we discuss the most likely evolutionary scenario which explains both the distribution and kinematics of the cold gas disc and the SFH of the quiescent and star-forming discs of NGC~1436, and in Sect.~\ref{sect:concl} we summarise our results. 

\section{Data}
\label{sec:observations}

\placefigPV

\subsection{MeerKAT}\label{sec:n1436meetkat}

We study the distribution and kinematics of \hi{} in NGC~1436 using data products from the MeerKAT Fornax Survey. The survey goals, design, observations and \hi{} data processing are described in \cite{2023arXiv230211895S}. Here we make use of the \hi{} subcubes and moment images of NGC~1436 at a resolution of $11$\arcsec and $66$\arcsec (see Table 2 in \citealp{2023arXiv230211895S}). The 11\arcsec cube has a noise level of 0.28~\mJb{} and a $3\sigma$ \hi{} column density sensitivity of $4.5\times10^{19}$~cm$^{-2}$.
The 66\arcsec cube has a noise level of 0.29~\mJb{} and a $3\sigma$ \hi{} column density sensitivity of $1.3\times10^{18}$~cm$^{-2}$. In both cases the column density sensitivity is calculated assuming a linewidth of 25~\kms. 
The spectral resolution of both cubes is 1.4~\kms.

\subsection{ALMA}\label{sec:n1436alma}

The ALMA observations of the $^{12}$CO(1–0) line used in this work, whose data reduction is described in \cite{zabel2021MNRAS.502.4723Z}, consist of a combination of high resolution data from the ALMA Fornax Cluster Survey obtained with the 12-m configuration array \citep{zabel} and deep data from the ALMA archive \citep{2022ApJS..263...40M} obtained with the Atacama Compact Array. 
The final cube has a velocity resolution of 10$~$\kms, a synthesized beam size of 2.68\arcsec$\times$2.06\arcsec, and a noise level of $\sim$2.0~\mJb{}. 


\subsection{MUSE}\label{sec:n1436muse}

The MUSE data used in this work were collected as part of the Fornax 3D survey \citep{sarzi2018A&A...616A.121S}.
The MUSE telescope \citep{2010SPIE.7735E..08B} ensures high-quality 3D spectroscopy with: (\emph{i}) a spatial sampling of 0.2\arcsec$\times$0.2\arcsec in a field of view of 1\arcmin$\times$1\arcmin, per pointing; (\emph{ii}) a wavelength range of 4650-9300$~$\r{A} with a nominal spectral resolution of 2.5$~$\r{A} (FWHM) at 7000$~$\r{A} and a spectral sampling of 1.25$~$\r{A}/pixel.
Due to the combination of different, slightly offset, exposures taken at different position angles (PAs), the average spectral resolution is 2.8$~$\r{A} (FHWM).
The data reduction, performed using the MUSE pipeline version 1.6.2 \citep{2012SPIE.8451E..0BW, 2016ascl.soft10004W} is described in \cite{sarzi2018A&A...616A.121S}. 

\section{Morphology and kinematics of the cold gas disc}\label{sec:n1436gas}

Since \hi{} discs are usually more extended than the stellar ones, they are excellent tracers of recent/on-going interactions with the local environment.
As mentioned in Sect.~\ref{sec:n1436intro}, previous results on the \hi{} emission in NGC~1436 \citep{loni21} are not exhaustive and also have a large discrepancy with single dish data from \cite{courtoisGBT}. The high sensitivity, spatial and spectral resolution of the MeerKAT Fornax Survey data fulfils well our needs, providing us with an ideal data-set to study in detail the properties of the cold gas disc of NGC~1436. The high-resolution cube allows us to gain information both on the \hi{} morphology and kinematics, while the low-resolution cube, with better column density sensitivity, might recover more flux.

Fig.~\ref{fig:mom0s} shows both the $11''$ and $66''$ resolution HI images, while the bottom left panel shows their comparison after smoothing the former to the resolution of the latter. Note that the 11\arcsec image was made with a different Briggs robust and a different detection mask than the 66\arcsec image, and therefore could in principle miss some of the 66\arcsec outer flux (see \citealp{2023arXiv230211895S}). We see that the match between these contours is excellent, meaning that NGC~1436 does not host any diffuse, low-column density \hi{} in excess of the \hi{} detected at high resolution. This implies that the \hi{} disc is sharply truncated and does not extend beyond the inner, star-forming disc of NGC~1436. Since the data are extremely sensitive, and the \hi{} cubes do not show any evidence of distortions, asymmetries or tails at low column density, we can confidently rule out any recent disturbance to the gas disc, while we see the presence of a \hi{} depression in the center of the galaxy. 

The agreement between the 11\arcsec~and 66\arcsec~\hi{} data cubes is also clear in the bottom right panel of Fig.~\ref{fig:mom0s}, where we compare the spectra extracted using the respective detection masks. 
We also show the single-dish spectrum (GBT - \citealp{courtoisGBT}), whose total flux is consistent with those measured with MeerKAT within 1$\sigma$.
Using Eq. 50 in \cite{Meyertracing}, we estimate the total \mhi{} to be $\left(1.9\pm0.2\right)~\times~10^8$~M$_\odot$.
Based on ATCA data, \cite{loni21} pointed out that NGC~1436 is a \hi{}-deficient galaxy with a large $M_{\mathrm{H_2}}$/\mhi{} ratio. These results do not change if we use the MeerKAT \mhi{} value, which is 4$\sigma$ larger than the ATCA one, the same \mst{} value calculated in \cite{loni21}, and the $M_{\mathrm{H_2}}$ value by \cite{zabel2021MNRAS.502.4723Z} (which is a factor $\sim$1.4 smaller than that used in \citealp{loni21} from \citealp{zabel}). Indeed, (\emph{i}) the offset between the \mhi{}/\mst{} ratio of NGC 1436 and the xGASS \mhi{}-\mst{} scaling relation \citep{BarbaraxGASS} is still larger than the RMS deviation of a sample of non-cluster galaxies \citep{VoidKreckel2012, HRSboselli2014} and (\emph{ii}) the $M_{\mathrm{H_2}}$/\mhi{} ratio of NGC~1436 is 1.0$\pm$0.1, resulting $\sim12~\sigma$ above the xGASS weighted average of $\log_{10}$($M_{\mathrm{H_2}}$/$M_{\mathrm{H_I}}$). 

The top left panel in Fig.~\ref{fig:HIH2comp} shows the $g$-band Fornax Deep Survey optical image \citep{FDSIodice2016,venhola2018fds,Reynier_2020arXiv200812633P} of NGC~1436 where we overlay both the $11''$ MeerKAT \hi{} image contours (red and black contours are the same as in the top right panel of Fig.~\ref{fig:mom0s}) and the \htwo{} contours (white color) from \cite{zabel2021MNRAS.502.4723Z}. Here, we can see that the \htwo{} follows the \hi{} ring and that there is no \htwo{} detected in the center of the galaxy. In the top right panel of the figure we show the \hi{} velocity field within the lowest reliable \hi{} contour, where we see that the \hi{} disc is regularly rotating with a slight PA warp at the edge of the disc. This is visible as an anticlockwise rotation of the outer isovelocity contours.
The bottom panels show the position-velocity diagrams (PVDs) along the kinematical major axis (left, PA = 330$^{\circ}$) and minor axis (right, PA = 240$^{\circ}$) of both \hi{} (background colour) and \htwo{} (contours). The rotational velocity of \hi{} and \htwo{} are consistent with one another and follow the typical rotation of a settled disc.
Judging from the appearance of the velocity field and the \hi{} PVD, we see that the rotation curve is rising in the inner $\sim$half of the \hi{} disc, and flattens further out. 

\placefigTir

In the rest of this section we show results obtained by fitting the \hi{} kinematics of the high-resolution cube with the Tilted Ring Fitting Code\footnote{The software is available at the following link \url{https://gigjozsa.github.io/tirific/index.html}} (TiRiFiC - \citealp{tirific2012ascl.soft08008J}). TiRiFiC models the galaxy in rings, where the model parameters are fitted to best reproduce the \hi{} geometry and kinematics in the data (further details in \citealp{tirific22007A&A...468..731J}). This allows us to measure the \hi{} surface brightness profile and to find the best-fitting values for geometric parameters (e.g. inclination, PA) and for the rotation curve. 
We followed two modelling approaches. In the former, we model our galaxy with a flat disc  (i.e., we forced TiRiFiC to fit a single value for PA and inclination across all rings). In the latter, we forced TiRiFiC to fit a flat rotation curve, but allowed the disc to warp letting the inclination vary. Between these two approaches, we obtained better results modelling NGC~1436 with a flat disc. We further improve the output flat disc model by using an advanced feature of TiRiFiC, which allows us to fit and interpolate some model parameters in groups of model rings rather than for each ring independently. Eventually, we chose the model that better follows the distribution and kinematics of the observed \hi{} based on visual inspection. Appendix~\ref{sec:app} includes a visual comparison between the data and the TiRiFiC model. 

Fig.~\ref{fig:SBR} shows the \hi{} rotation curve and the surface brightness (SB) of NGC~1436 obtained through our TiRiFiC fitting. As anticipated, the velocity increases with galactocentric distance in the inner regions, and it reaches a plateau at $\sim$50\arcsec.
In the bottom panel, we show the derived SB, which peaks at 45\arcsec. The best-fitting model shows some low level fluctuations, which are unlikely to be real and disappear when convolving the profile with the 11\arcsec~beam (orange line in Fig.~\ref{fig:SBR}). It is worth noting that, if we use our best model as first guess and let PA and inclination vary, we obtain the same SB and rotational velocity.

The \hi{} disc diameter of NGC 1436, measured where the surface brightness is 1~M$_\odot~$pc$^{-2}$, is $\sim$110\arcsec ($\sim11$~kpc). This allows us to verify that NGC~1436 is 2.26$\sigma$ above - and thus in agreement with - the \hi{} size - mass relation \citep{wang2016_size} similarly to the truncated \hi{} discs in the Virgo cluster.

In general, by studying the distribution and kinematics of both \hi{} and \htwo{}, we appreciate the high level of symmetry of the cold gas in NGC~1436. Both \hi{} and \htwo{} are distributed within a settled ring where they are co-rotating, with no hints of any recent tidal or hydrodynamical disturbances. Given that the orbital time at the disc's edge is of $\sim$300~Myr (as derived from the de-projected rotational velocity of $\sim$130~\kms at a radius of $\sim~$6.7~kpc) we can rule out strong interactions in the last few orbital times, roughly 1~Gyr.

\section{Star formation history}\label{sec:n1436sfhresult}

The analysis on the cold gas disc establishes that previous interactions between the local environment and NGC~1436 left the galaxy with a relaxed truncated \hi{} disc and a clear demarcation between the star-forming inner disc regions and passively-evolving outer ones. In this section, we now aim at reconstructing the history of this morphological transition by tracing the SFH imprinted in the inner and outer stellar populations.
For this we adopt a non-parametric approach based on single-age stellar population models (SSP) and on the use of the Penalised Pixel-Fitting \footnote{\url{https://www-astro.physics.ox.ac.uk/~mxc/software/}} (pPXF; \citealp{ppxf2004PASP..116..138C}) and the Gas and Absorption Line Fitting codes\footnote{\url{https://star.herts.ac.uk/~sarzi/}} (GandALF; \citealp{gandalf2006MNRAS.366.1151S,gandalf22006MNRAS.369..529F}) to fit the MUSE data.  
However, rather than aiming at a regularized SFH solution \citep{ppxf22017MNRAS.466..798C} we decide to exploit the axisymmetry of the system to capture the uncertainties of the SFH-reconstruction, as done earlier using MUSE data by \cite{2018Martinsson}. 

\placefigMUSEHalphaapp

More specifically, we derive the SFH for two sets of spectra extracted from sectors of elliptical annuli covering the star-forming and quiescent disc, respectively, as shown in Fig.~\ref{fig:SFH_regions}.     
Since the stellar populations within such annular sectors should be rather similar for the inner and outer disc, respectively, beyond the short timescales ($\lesssim\! 100$ Myr) corresponding to the orbital times between adjacent sectors, we use the variance between the SFHs recovered in such sectors to estimate the uncertainty. In total we define 18 sectors in the inner disc and 8 sectors in the outer disc, with  average S/N per pixel of 100 and 20, respectively. Where necessary, foreground and background sources were masked out before extracting these aperture spectra, which we also corrected for foreground Galactic extinction using the Calzetti dust extinction law \citep{2000ApJ...533..682C} and a $A(V) = 0.029$ value in the direction of Fornax from \cite{2011ApJ...737..103S}.

To derive the SFH from each of these aperture spectra, we fit them with pPXF and GandALF using SSP models from the MILES library of \cite{2015MNRAS.449.1177V}\footnote{The MILES stellar-population model library can be found at the following link \url{http://miles.iac.es/}}, which covers a range in age between 30~Myr and 14~Gyr, with total metallicity $-2.27\leq$[M/H]$\leq0.4$~dex, and $\alpha$-element overabundance [$\alpha$/Fe] between 0 and 0.4~dex. Since without a dedicated methodology (e.g. \citealp[]{2019A&A...626A.124M, 2021A&A...654A..59M}) we do not expect to be able to capture variations in the stellar initial mass faction (IMF) through direct spectral fitting, here we further restrict the fits to using models with a Kroupa IMF \citep{2001MNRAS.322..231K}.

\placefigspe

\placefigsinglesfh
As in \cite{sarzi2018A&A...616A.121S}, we first fit the MUSE data with pPXF to measure the stellar kinematics using only the bluer part of the MUSE spectrum in the 4800 -- 5850~\r{A} spectral range. Subsequently, we extend the GandALF fitting range out to 6850~\r{A} to also include the typically strong H$\alpha$, [{\sc N$\,$ii}],
and [{\sc S$\,$ii}] emission and in turn better constrain the overall contribution of the emission-line spectrum to our models.
During the pPXF fit we mask regions potentially affected by emission and adopt an additive correction for the shape of the continuum. Using a shorter wavelength range speeds up the pPXF fitting process and delivers robust kinematic results owning to the presence of several strong stellar absorption features at the bluer end of the MUSE spectra. 
The GandALF fit is constrained to the pPXF kinematics, but is allowed to change the optimal combination of SSP templates in the stellar models in response to the additional use of emission lines in the fit, which are represented by Gaussian functions.
In particular, we include both the H$\beta$ and H$\alpha$ recombination lines and the [{\sc O$\,$iii}]$\lambda\lambda4959,5007$, [{\sc N$\,$i}]$\lambda\lambda5198,5200$, [{\sc O$\,$ii}]$\lambda\lambda6300,6363$,
[{\sc N$\,$ii}]$\lambda\lambda6548,6583$, [{\sc S$\,$ii}]$\lambda\lambda6713,6730$ forbidden ones, tying all lines to the same, optimally-adjusted line profile, while also accounting for the varying spectral resolution of the MUSE spectra \citep[e.g.,][]{Guerou16}.

Ideally, the GandALF fit would include a two-component reddening correction: the first affecting the whole spectrum to mimic diffuse dust; and the second affecting the emission line spectrum only, representing dust localised in emission line regions and matching the observed Balmer decrement (\citealp[e.g.,][]{OSSY}). 
However, the quality of the relative flux calibration of the MUSE data is insufficient to grant such an approach, and we resort to using a multiplicative polynomial correction instead.
This still allows us to obtain very good fits to our aperture spectra, but the use of multiplicative polynomials effectively introduces a nuisance parameter in our analysis as polynomial correction of varying order will lead to fits of very similar quality beyond a certain minimum order. 
Furthermore, not being able to account for dust reddening limits our control over unrealistic values for the Balmer decrement and spurious H$\beta$ line detections, which in turn can affect the stellar-template optimal combination process and lead to biased SFH results.     
For these reasons we decide to omit the H$\beta$ line from our GandALF fits for the outer quiescent disc, where H$\alpha$ emission from diffuse ionised gas is only barely detected.
Finally, we note that in the fits we further ignore the wavelength region around the NaD$\lambda\lambda 5896,5890$, as often this part of the spectrum presents strong continuum shape issues related to the relative flux calibration.

\placefigfinalsfh

Figure~\ref{fig:MUSEdatapoly} shows examples of aperture spectra extracted in the outer quiescent disc (left) and in the inner star-forming disc (right) of NGC~1436, along with our best GandALF fit to them. Residuals from the sky emission-line subtraction process are evident in the outer-disc aperture spectra but do not affect the overall quality of our fit nor our SFH results, in particular as these artifacts do not interest the more age- and metallicity-sensitive absorption-line features in our fitting range. 
The fits in Fig.~\ref{fig:MUSEdatapoly} include a 7-th order multiplicative polynomial adjustment that is more substantial in the case of the central star-forming regions, where such a correction is also needed to mimic the impact of reddening by dust.

During the fitting process, GandALF finds the best combination of our adopted stellar population templates returning a corresponding list of best-fitting weights. Since each MILES model corresponds to a stellar population with an initial mass of one solar mass, accounting for the distance to the object allows us to turn each of these weights into a stellar mass formed at the look back time corresponding to the age of the $i$-th SSP model. Given that only a few SSP templates receive a non-zero weight during the GandALF fit, to estimate the SFH in each sector $s$ we bin the resulting initial stellar-mass values $M_{i,s}^\mathrm{SSP}$ in 1-Gyr-wide age bins between 0 and 6~Gyr and in 2-Gyr-wide bins beyond that. Dividing the total stellar mass $M_s$($t_j$) formed in the $j$-th time bin by the corresponding time bin width $\Delta t_j$, we obtain the average star-formation rate (SFR) in that bin $\mathrm{\emph{SFR}}_s$($t_j$) = $M_s$($t_j$)/$\Delta t_j$. This effectively gives us a SFR history for each sector, which we divide by the total stellar mass formed over time, $M_s $=$ \Sigma_j M_s$($t_j$) to derive the fractional stellar mass formed per unit time in each sector. 

This normalised SFH varies strongly from sector to sector in both the inner disc and the outer disc. This variation largely reflects the uncertainties in our template-optimisation process. To estimate these uncertainties and obtain a more reliable average trend we calculate the average and standard deviation of the normalised star formation histories across all sectors assigning $SFR_s$($t_j$)~=~0 to sectors and age bins with no non-zero weights. Fig.~\ref{fig:SFHs_oneexp} shows the normalised SFHs for the inner and outer disc obtained after this averaging, using a 7-th order multiplicative correction as in Fig.~\ref{fig:MUSEdatapoly}.
The average values and corresponding uncertainties on the normalised star-formation history (coloured large circles and shaded regions, respectively) already indicate that both central and outer regions went through a recent peak of star-formation some 4 Gyrs ago.  event was much more pronounced in the inner disc (by a factor 4-5) and more rapidly quenched in the outer disc. 

However, we note that in the quiescent outer disc our GandALF fit seems to require a small contribution from very young SSPs. The presence of such young stars is at odds with the notable absence of any star-formation activity and it is certainly a bias in the template mix adjustment process, which in turn may stem from intrinsic limitations in our ability to constrain the presence of young stellar populations using the MUSE spectra, as indeed these do not extend down to important features (e.g. the D4000 break and H$\delta$).
For this reason we decided to repeat our analysis while also removing all SSP templates younger than 1~Gyr and include this second set of fits in the overall averaging and variance estimation process for the normalised SFHs. We highlight that the actual choice of polynomial order is also quite arbitrary, as the quality of the fit does not change significantly while varying the polynomial order between 6 and 15. Treating this as an additional source of uncertainty we thus repeat our fits varying the polynomial order within the above limits, including the corresponding results in the derivation of the final normalised SFH and associated uncertainties for the inner and outer disc.

\placeTNGbothsamecap 

Fig.~\ref{fig:SFHs_allexp} shows the final normalised SFH in the outer quiescent disc (left) and inner star-forming disc (right) obtained combining the normalised SFHs from their corresponding apertures, considering 20 different fits (10 polynomial orders, and including or excluding SSP templates younger than 1 Gyr). 
The clearer trends of Fig.~\ref{fig:SFHs_allexp} confirm the picture drawn from our simpler analysis shown in Fig.~\ref{fig:SFHs_oneexp}, according to which NGC~1436 experienced a burst of star formation between 3 to 4 Gyr ago that saw the birth of $\sim$50\% and $\sim$20\% of all the stars that ever formed in the inner and outer disc, respectively.
Such an activity faded considerably in the past 3 Gyr, and most dramatically in the outer disc where the absence of any present star-formation activity (from the H$\alpha$ data) is consistent with the lack of star formation between 1 and 2 Gyr in our reconstructed SFH.
Clearly, our ability to assess the rate of recent decline in star formation in NGC~1436 is limited by the aforementioned biases in recovering the SFH in the last 1 Gyr. 
As a sanity check, in the inner disc we can at least provide an estimate for the fractional stellar mass formed in the last Gyr assuming that stars formed at their present-day level over this period of time. 
Using the MUSE H$\alpha$ maps of \citet{IodiceF3D} and restricting our analysis to emission-line regions firmly classified as {\sc H$\,$ii} regions we obtained the total star-formation rate in the regions subtended by our inner-disc sectors, using the  \cite{2012ApJ...752...98C} conversion SFR(M$_\odot$/yr)~=~5.5$\times$10$^{-42}$~L$_{\mathrm{H}\alpha}~\rm(erg/s)$ and taking a distance of 20 Mpc.
The estimate derived in this way (blue cross in Fig.~\ref{fig:SFHs_allexp}) is consistent with the upper edge for the fractional stellar mass formed in the last 1 Gyr. Similarly, from the H$\alpha$ map of the outer disc, we measure a 3$\sigma$ SFR upper limit (red downward arrow) using the 8 sectors of the outer disc.
Finally, the reconstructed SFH of NGC~1436 shows that before the onset of its recent starburst, some 4 to 5 Gyrs ago, this disc galaxy was evolving according to the cosmological picture with its star-formation activity having already peaked at a redshift between 1 and 2, or some 8 to 10 Gyrs ago \citep{2018A&A...615A..27L}. In fact, the steeper decline in star-formation activity in the inner regions at these earlier times, is consistent with the commonly observed inside-out evolution of disc galaxies \citep[e.g.,][]{Kepner99,Nelson16, Johnston22}.   
In summary, our SFH analysis indicates that both the inner and outer disc populations went through three main star-formation phases: \emph{(i)} a rather typical build up activity and a steady decline in star-formation activity until some 5~Gyr ago, \emph{(ii)} a burst of star formation peaking between 3 and 4 Gyr ago, and \emph{(iii)} a final drop in star-formation until present times, gradually reducing it by two orders of magnitude in the inner disc and completely quenching star formation in the outer disc. 

\section{Discussion}\label{sec:n1436discussion}

Several properties of the multi-phase gas disc indicate that NGC 1436 has not experienced any significant tidal or hydrodynamical interaction in recent times: (\emph{i}) the highly symmetric morphology and settled kinematics of the \hi{} disc observed with MeerKAT at a resolution of $\sim 10$\arcsec $\sim 1$~kpc; (\emph{ii}) the absence of \hi{} disturbances, asymmetries or tails down to a column density of $\sim10^{18}$ cm$^{-2}$  (resolution $\sim 60$\arcsec$\sim 6$~kpc); \emph{(iii)} the perfect match between the regular morphology and kinematics of the various gas phases (atomic, molecular and ionised). The fact that the disc is settled now, does not imply that NGC~1436 has never experienced environmental interactions in the past. For example, in the Virgo cluster, some of the most gas-poor galaxies which went through ram-pressure stripping are symmetric systems with a severely truncated \hi{} disc \citep{2017ApJ...838...81Y}.
That is, after a few orbital periods any sign of interactions would disappear. Given the orbital time of NGC~1436 (300 Myr at the edge of the \hi{} disc - Sect.~\ref{sec:n1436gas}), NGC~1436 has not undergone any significant environmental interaction in the last few orbits, say $\sim1$ Gyr.
Thus we wonder "whether", "when" and "how" interactions with the Fornax cluster environment were relevant for the history of NGC 1436.


Useful insights come from our reconstructed SFH for the inner and outer disc of this galaxy. The SFH of the quiescent and star-forming disc of Fig.~\ref{fig:SFHs_allexp} indeed suggests that NGC~1436 was not a member of the cluster earlier than $\sim$6 Gyr ago, with the Fornax environment becoming relevant for its evolution starting around 5~Gyr ago. 
Indeed, between 14 and 5 Gyr NGC~1436 seem to have followed the expected cosmic evolution for star-formation activity  \citep{2018A&A...615A..27L} while also following a typical inside-out evolution. 
That is, up to $\sim5$ Gyr ago NGC~1436 seems to have formed its stars as a regular spiral in a low-density environment, without being affected by strong tidal or hydrodynamical interactions. 
Since entering the cluster it must have taken $\sim 1$ Gyr for NGC~1436 to reach its first pericentre passage ($\tau_\mathrm{cross}$ is indeed about 2 Gyr for Fornax where $R_\mathrm{vir}~$=$~700$~kpc and $\sigma_\mathrm{cl}~$=$~318$~\kms, \citealp[]{drinkwater2001substructure}) when interactions between the galaxy and the Fornax environment would have been more extreme.

Drawing from the inferred rise and decline of star formation activity, we thus speculate that $\sim4$ Gyr ago, at about pericentre, a tidal interaction and/or ram-pressure was able to compress and then remove the majority of the \hi{} content from the galaxy outer disc, causing a temporary SFR enhancement followed by a rapid quenching of its star formation. NGC 1436 could only keep some \hi{} within the inner star forming disc thanks to the stronger gravitational binding. However, by the time gas removal from the outer disc was complete, the accretion of cold gas onto the galaxy must have completely stopped. Therefore, the subsequent evolution (with NGC~1436 now moving away from pericentre) must have been characterised by a fully quenched outer disc and a starving inner disc. This would explain why the SFR of the inner disc has decreased in the last few Gyr, but not gone to zero. The lookback time of the environmental interactions, which caused the enhancement of the SFR, is consistent with the conclusion of \cite{IodiceF3D} that NGC 1436 is an intermediate infaller in Fornax.

The relative role that gravitational and hydrodynamical processes might have played to drive such a morphological transformation is not clear.
In the literature there is evidence that both tidal interactions and ram pressure might separately trigger a temporary enhancement in star-formation activity and the removal of cold gas \citep{1972ApJ...178..623T,2004MNRAS.350..798B,2008MNRAS.385.1903L,2017ApJ...844...48P,2018ApJ...866L..25V,Mpati2019MNRAS.487.4580R, 2022A&A...659A..94M}. Alternatively, a combination of both gravitational (possibly a merger) and hydrodynamical interactions might have been at work in the specific case of NGC~1436. For example, NGC~1436 might have experienced tidal interaction near pericentre, which may have triggered a central SFR enhancement while, at the same time, moving some gas to large radius via tidal forces, from where it could more easily be stripped by ram pressure. In this case, NGC~1436 would share a similar evolutionary path to that of the Fornax galaxies with one-sided HI tails described in \cite{2023arXiv230211895S}. 

Interestingly, the recent evolution of NGC~1436 can also be observed in cosmological simulations. For instance, looking at the TNG50 simulation \citep{Pillepich19}, among the 16 present-day disc galaxy found in Fornax-like environments \citep{PabloTNG50} we find six currently star-forming objects that entered their cluster environment between 3 and 5 Gyrs ago and had only one pericentre passage. 
Among these, the left panel of Fig~\ref{fig:sfr_tng} shows the SFH of the simulated galaxy that matched more closely the one inferred from our SFH analysis. The right panel of this figure shows that the disc of this simulated object also appears to be  forming stars only in its inner part while its outskirts have already been quenched, as for NGC~1436.
\section{Summary}\label{sect:concl}

Using \hi{} cubes from the MeerKAT Fornax Survey \citep{2023arXiv230211895S}, with a resolution of 11\arcsec~and 66\arcsec, sensitivity $N_{\mathrm{HI}}$($3\sigma,~25~\mathrm{km s^{-1}}$)~=~$4.5\times$10$^{19}$~cm$^{-2}$ and $1.3\times$10$^{18}$~cm$^{-2}$, respectively, and a spectral resolution of $\sim$1.4\kms, we have investigated the distribution and kinematics of the \hi{} in NGC~1436 finding that: \emph{(i)} the \hi{} distribution is truncated within the inner and still star-forming disc \emph{(ii)} the \hi{} kinematics is consistent with that of \htwo{} \citep{zabel} and ionized gas \citep{IodiceF3D}, and thus NGC~1436 has not experience any recent environmental interactions in the last $\sim$1~Gyr (a few orbital times of the interstellar medium). 

With Fornax~3D data \citep{sarzi2018A&A...616A.121S}, we further reconstructed the SFHs of the outer (currently quiescent) and inner (still star-forming) part of the disc of NGC~1436, which show (\emph{i}) a faster evolution of the inner disc with respect to the outer disc from 14 to 5 Gyr ago (\emph{ii}) a burst of star formation $\sim4$ ago in both parts of the disc (\emph{iii}) a steady decline and a fast quenching of the star formation activity in the inner and outer disc, respectively.

These findings suggest that before the star formation burst, NGC~1436 was evolving in a low-density environment outside the Fornax cluster. While falling into the cluster and approaching its pericentre, environmental interactions have promoted a temporary enhancement in SFR. These interactions were also able to remove \hi{} from the outskirts of the galaxy, which caused the fast quenching of the outer disc. The lack of subsequent cold gas accretion onto the inner disc is causing its current slow quenching.

Similar evolutionary paths have been observed in cosmological simulations \citep{Pillepich19} in disc galaxies evolving in Fornax-like clusters \citep{PabloTNG50}.  

\section{Future plan}\label{sect:future}

This work demonstrates how mapping the star-formation history in galaxies with deep \hi{} data is effective
for understanding both the past and on-going impact of gravitational and hydrodynamic processes on galaxies moving from low to high density environments.

As a next step, we will apply the stellar-population analysis approach used for NGC~1436 to the other galaxies in the Fornax cluster with available VLT-MUSE data and detected with MeerKAT. Mainly coming from Fornax3D sample (\citealp{sarzi2018A&A...616A.121S}), they will result in a sample of 13 galaxies covering a range of HI morphologies, galaxy mass, and position within the cluster that will provide an ideal basis for further comparisons with the predictions from numerical simulations.
At the same time, we intend to improve our constraints on the young stellar population by taking also into account data extending beyond the MUSE spectral range. For instance, we intend to collect near-IR data using the NIRWALS integral-field spectrograph \citep{2018SPIE10702E..2OW} mounted on the South African Large Telescope (SALT), which will cover absorption bands from thermally-pulsating asymptotic giant branch stars that are good indicators of intermediate age populations (between 0.3 and 2 Gyr, \citealp{2005MNRAS.362..799M}).


\section*{Acknowledgements}

This project has received funding from the European Research Council (ERC) under the European Union’s Horizon 2020 research and innovation programme (grant agreement no. 679627; project name FORNAX). The MeerKAT telescope is operated by the South African Radio Astronomy Observatory, which is a facility of the National Research Foundation, an agency of the Department of Science and Innovation. We acknowledge the use of the Ilifu cloud computing facility - www.ilifu.ac.za, a partnership between the University of Cape Town, the University of the Western Cape, the University of Stellenbosch, Sol Plaatje University, the Cape Peninsula University of Technology and the South African Radio Astronomy Observatory. The Ilifu facility is supported by contributions from the Inter-University Institute for Data Intensive Astronomy (IDIA - a partnership between the University of Cape Town, the University of Pretoria and the University of the Western Cape), the Computational Biology division at UCT and the Data Intensive Research Initiative of South Africa (DIRISA). This project and both AL and MS have received support from the Science and Technologies Facilities Council (STFC) through their New Apllicant grant ST/T000503/1.  
FL acknowledges financial support from the Italian Minister for Research and Education (MIUR), project FARE, project code R16PR59747, project name FORNAX$-$B. 
FL acknowledges financial support from the Italian Ministry of University and Research -- Project Proposal CIR01$\_$00010.
J.~F-B  acknowledges support through the RAVET project by the grant PID2019-107427GB-C32 from the Spanish Ministry of Science, Innovation and Universities (MCIU), and through the IAC project TRACES which is partially supported through the state budget and the regional budget of the Consejer\'ia de Econom\'ia, Industria, Comercio y Conocimiento of the Canary Islands Autonomous Community. 
EMC acknowledges support by Padua University grants DOR 2020–2022 and by Italian Ministry for Education University and Research (MIUR) grant PRIN 2017 20173ML3WW-001. AP is supported by the Science and Technology Facilities Council through the Durham Astronomy Consolidated Grant 2020–2023 (ST/T000244/1). This project has received funding from the European Research Council (ERC) under the European Union's Horizon 2020 research and innovation program grant agreement No. 882793, project name MeerGas. 
NZ is supported by the South African Research Chairs Initiative of the Department of Science and Technology and National Research Foundation. At RUB, this research is supported by the BMBF project 05A20PC4 for D-MeerKAT.

\section*{Data Availability}

The data underlying this article from: \emph{(i)} the MeerKAT Fornax Survey are available at the following link \url{https://sites.google.com/inaf.it/meerkatfornaxsurvey/data}, \emph{(ii)} the Fornax3D survey are available in the ESO archive (program ID: 296.B-5054 - see also \href{http://archive.eso.org/cms/eso-archive-news/new-data-release-of-muse-data-cubes-of-the-fornax3d-survey.html}{new-data-release-of-muse-data-cubes-of-the-fornax3d-survey}), \emph{(iii)} the ALMA Fornax Cluster Survey and the Deep CO(J=1-0) mapping survey of Fornax galaxies with Morita array are available in the ALMA archive (project code 2015.1.00497.S and 2017.1.00129.S, respectively).
 



\bibliographystyle{mnras}

\bibliography{biblio2} 



\clearpage
\appendix
\section{}\label{sec:app}

\begin{figure*}
    \centering
    \includegraphics[width=0.90\linewidth]{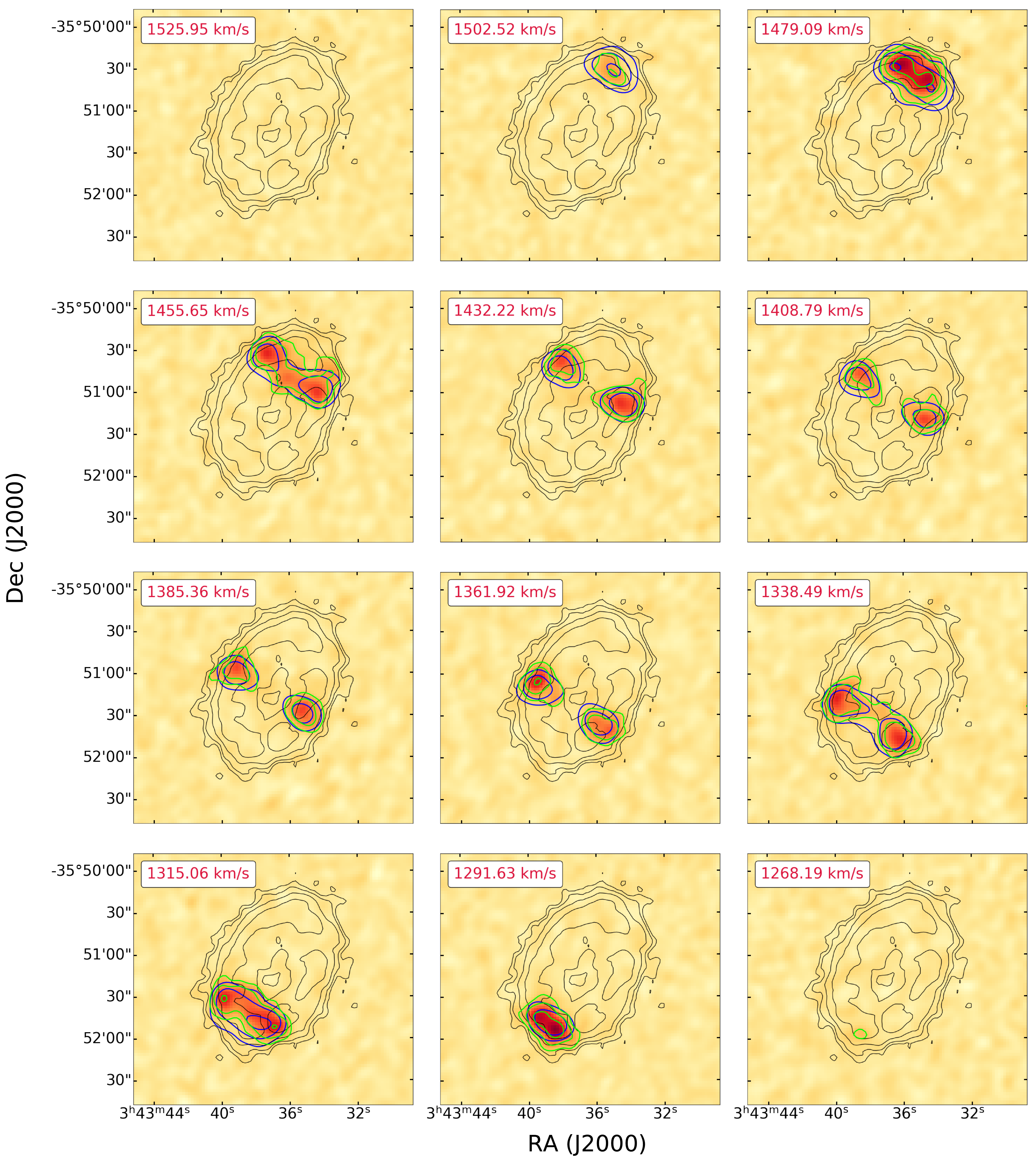} 
    \caption{We show the comparison between the channel-maps of the 11\arcsec cube (green) and the TiRiFiC model (blue) used to measure the rotational velocity and SB in Fig.~\ref{fig:SBR}. For presentation purpose, we re-binned the channel-maps in velocity to 23~\kms~channels. We show the same contours at $\sim$3$\sigma$ (where $\sigma$ is the RMS of the re-binned cube), with subsequent contours increasing as $\times$2$^n$ (with n=1,2,3 ..), both for the data and the model. In general, the model follows the \hi{} rotation in the data, and in the two middle rows, we can see how TiRiFiC nicely models the \hi{} depression visible in the center of the galaxy. Black contours are the 11\arcsec \hi{} image (Fig.~\ref{fig:mom0s}).}\label{fig:chanmap_model}
\end{figure*}



\bsp	
\label{lastpage}
\end{document}